\documentclass[%
reprint,
superscriptaddress,
 amsmath,amssymb, 
 onecolumn, 
]{revtex4-2}

\usepackage{graphicx,xcolor}
\usepackage{dcolumn}
\usepackage{bm}
\usepackage[version=4]{mhchem}
\usepackage{scalerel}
\usepackage{mathtools,hhline,array}
\newcolumntype{P}[1]{>{\centering\arraybackslash}p{#1}}
\allowdisplaybreaks

\begin{document}


\title{Diffusiophoresis in the presence of a pH gradient}

\author{Suin Shim}
\email{sshim@princeton.edu}
 \affiliation{Department of Mechanical and Aerospace Engineering, Princeton University, Princeton, NJ 08544}
\author{Janine K. Nunes}%
\affiliation{Department of Mechanical and Aerospace Engineering, Princeton University, Princeton, NJ 08544}%

\author{Guang Chen}
\affiliation{Department of Advanced Manufacturing and Robotics, College of Engineering, Peking University}


\author{Howard A. Stone}
\email{hastone@princeton.edu}
\affiliation{Department of Mechanical and Aerospace Engineering, Princeton University, Princeton, NJ 08544}%

\date{\today}

\begin{abstract}
\vspace{1ex}
Diffusiophoresis is the spontaneous motion of particles under gradients of solutes. In electrolyte-driven diffusiophoresis, the zeta potential of the particles is an important surface property that characterizes diffusiophoretic mobility. However, the zeta potential is not a fixed material property and colloidal surfaces often show varying potentials depending on the physical chemistry of the surrounding fluid, e.g., solute type, ionic strength, and pH. In this article, we study experimentally and theoretically pH-dependent diffusiophoresis of polystyrene particles  using a dead-end pore geometry. In particular, the influence of the isoelectric point (pI) on diffusiophoresis is demonstrated in the absence and presence of wall diffusioosmosis. Throughout the paper, we show with experiments and model calculations how the pH-dependent diffusiophoresis and diffusioosmosis influence the particle motion in dead-end pore configurations, including changes that occur when there is a sign change in the zeta potential near the pI. 
\end{abstract}

\maketitle


\section{Introduction}

Diffusiophoresis is the motion of charged particles under solute concentration gradients, where the particle velocity is dependent on various electrokinetic properties at the surface \cite{derjaguin1947, derjaguin1947E,derjaguin1961, derjaguin1961E, dukhin1982, anderson1984,prieve1984jfm, anderson1984, ebel1988,staffeld1988-1,anderson1989review,dukhin1993review}. When the solute is an electrolyte, the zeta potential, which is an equilibrium potential at a shear plane in the diffuse double layer, is an important surface property that determines the magnitude and direction of the particle motion. The zeta potential is a theoretical value that is defined in the liquid phase in contact with a surface. Unless the surface is a constant-potential material, the zeta potential is usually not a fixed number and varies depending on the physical chemistry, i.e., the ionic strength, pH, solute type, etc., of the surrounding liquid \cite{kirby2004-1,kirby2004-2}. When the pH of a liquid phase is considered, adsorption or binding of \ce{H+} ions on a particle surface appears as variations in the surface charge density and the corresponding zeta potential in the electrical double layer (EDL) \cite{uematsu, zimmermann1, zimmermann2, healy1978ionizable, healy1980, healy1985, healy1992, saville1992}. 

In this article, we are particularly interested in the influence of \ce{H+} concentration on diffusiophoresis of polystyrene microspheres (diameter $\approx 1~\mu$m). The influence of ionic strength on the diffusiophoretic mobility for the case where ions do not react with surface functional groups is studied in \cite{gupta2020sm}. Polymeric particles are often assumed to have a charge regulation surface, where the surface charge density is controlled by the extent of proton binding at specific sites. Depending on how the surface is formulated chemically, such colloidal microspheres can have zeta potentials that are different functions of ionic strength, pH, dielectric perimittivity, solute type, etc. \cite{kirby2004-1,kirby2004-2,healy1978ionizable, healy1992, saville1992}. Commercial polystyrene (PS) particles that are used commonly in experimental studies of diffusiophoresis have a negative surface potential, with no isoelectric point (pI; the pH value where the zeta potential is zero) \cite{squires2016}. Amine-modified polystyrene (a-PS) particles typically show a positive surface potential \cite{apszeta1, apszeta2, shim2020pulse, shim2021hsc, shim2021wc}. To the best of our knowledge, the influence of a pH-dependent zeta potential has not been reported in the context of diffusiophoretic-driven particle motion caused by a pH gradient. At a moderate pH, amine functional groups on the particle surface bind with \ce{H+} to  form \ce{NH3+}, but, similar to common proteins \cite{gelelectrophoresis, proteinpurification, BSA, lysozyme}, it is likely that the concentration of \ce{H+}-bound surface groups decreases as the pH increases. If such particles have an isoelectric point, then the diffusiophoretic mobility will be affected by the sign change in the zeta potential. Thus, this feature suggests that in situations where the electrophoretic contribution to diffusiophoresis dominates chemiphoresis, the direction of particle motion in a concentration gradient can be flipped near the pI in the presence of a pH gradient.

We are not the first group to report diffusiophoresis of polystyrene particles in the systems with a pH gradient \cite{squires2016,steepphgradient,lee2018,seo2020}. Previous studies that used acidic and basic solutions or a Nafion membrane discuss the pH change in the system. However, the existing studies do not include particles that show a dramatic change in surface potentials or have an isoelectric point, and thus explanations for diffusiophoretic mobilities do not include zeta potential as a function of pH. Theoretically, diffusiophoresis of charge-regulating particles has been investigated for various types of particles (soft, porous, polyelectrolyte, etc.) \cite{hsu2015,crdiff1,crdiff2,crdiff3,crdiff4}. The studies do not discuss diffusiophoretic motion of charge-regulating particles under a pH gradient. Rather, they report diffusiophoresis under a KCl or NaCl gradient, at different fixed pH values. Our main argument focuses on the situations where a pH-dependent zeta potential and the existence of the surface pI are important in the analyses of diffusiophoresis in the presence of a pH gradient. 

We motivate our study with a set of compaction experiments \cite{wilson2020, alessio2021prf} in a dead-end pore geometry. Under the concentration gradients of HCl and NaOH (set up separately), polystyrene (PS) and amine-modified polystyrene (a-PS) particles in the pores move toward the dead-end by diffusiophoresis. When a-PS particles are initially suspended in 10 mM NaOH, we observe that the particles (originally positively charged at lower pH) behave like negatively charged PS. In Sections \ref{SectionII}-\ref{SectionIV}, by showing zeta potentiometry data, we rationalize this unexpected observation in the a-PS diffusiophoresis experiments. The zeta potential measurements are then fitted by a charge regulation model considering both acidic and basic functional groups. In order to systematically study the diffusiophoresis of a-PS particles under pH gradients we design a set of dead-end pore experiments that show a finite penetration of a patch of particles. The design naturally sets up a pore environment where the influence of wall diffusioosmosis can be neglected. Therefore, by combining charge regulation and  multi-ion diffusiophoresis models, we predict one-dimensional (1D) particle trajectories describing the time evolution of a front of particles penetrating into a pore.  

In the following section (Section \ref{SectionV}), we show with experiments and model calculations the situations where inclusion of diffusioosmosis at the channel walls is necessary. PDMS (polydimethylsiloxane) walls are highly negatively charged at high pH, and thus the presence of NaOH concentration gradients affects the particle motion by the combined pH-dependent diffusiophoresis and diffusioosmotically-driven liquid flow. Multi-ion diffusiophoresis calculations suggest that neglecting diffusioosmosis in some cases can lead to misinterpretation of the one-dimensional (1D) particle motion. Finally, in Section \ref{SectionVI}, we show model calculations for the diffusiophoresis of particles that have different pIs. Three different pH gradients are used: no pH gradient, $2\le \hbox{pH} \le 7$, and $7\le \hbox{pH} \le 12$. As diffusiophoresis of biological particles \cite{shim2021hsc,ramm2021,protein2012,protein2020rs,fahim2020, bloodcell2018, rasmussen2020nc} is of increasing interest in the research community, our systematic study of several model scenarios can provide the basis for new insights into the role of the chemical environment on the dynamics of natural and complex systems.

\section{Diffusiophoresis of charged particles in the presence of a pH gradient}
\label{SectionII}

\begin{figure}[t!]
\centering
\includegraphics[width=0.953\textwidth]{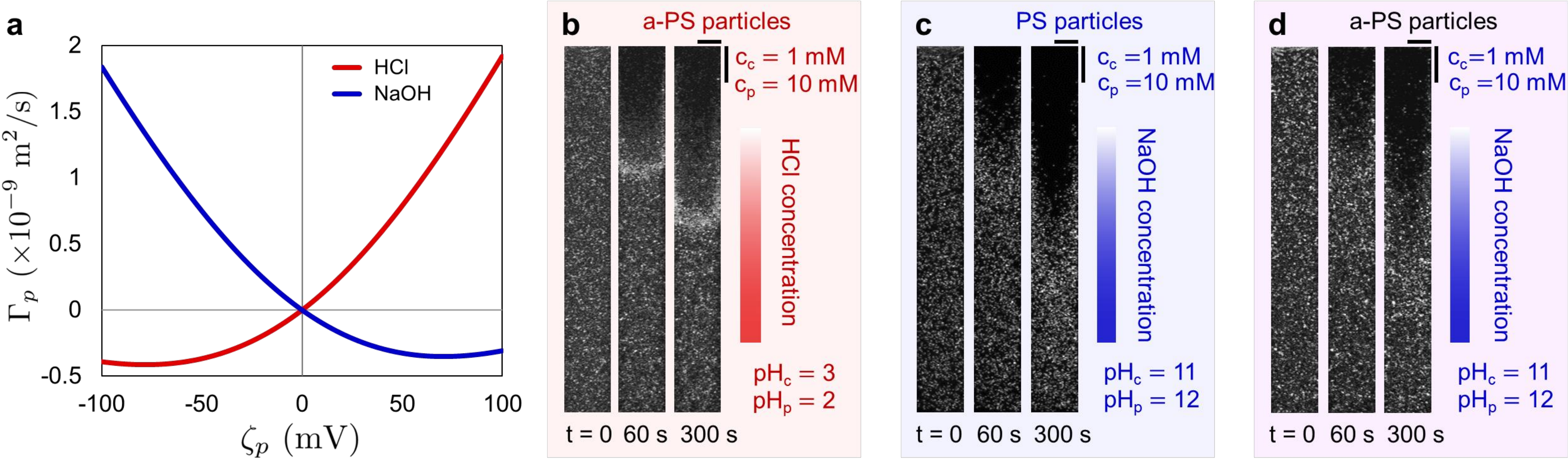}
\caption{\label{fig1} Diffusiophoresis of polystyrene particles under concentration gradients of \ce{HCl} and \ce{NaOH}. (a) Diffusiophoretic mobility (equation 1) for \ce{HCl} and \ce{NaOH} solutions plotted versus zeta potential ($\zeta_p$). The diffusivity difference factors are, $\beta_{\ce{HCl}}=0.642$ and $\beta_{\ce{NaOH}}=-0.596$ (see Table 1). Within the plotted range, the sign of mobilities change as the sign of zeta potential changes, indicating that the electrophoretic contribution of diffusiophoresis is dominant. Positively charged particle move up the concentration gradient of \ce{HCl} and negatively charged particles move up the concentration gradient of \ce{NaOH}. (b) Compaction of amine-modified polystyrene (a-PS) particles along an \ce{HCl} concentration gradient ($c_p=10$ mM and $c_c=1$ mM; pH$_p=2$ and pH$_c$=3). (c,d) Compaction of (c) polystyrene (PS) and (d) a-PS particles along a \ce{NaOH} concentration gradient (Video S1; $c_p=10$ mM and $c_c=1$ mM; pH$_p=12$ and pH$_c$=11). (d) We observe that at high pH (= 11-12), a-PS particles behave like negatively charged particles (compact into the pore). (b-d) Horizontal and vertical scale bars are, respectively, 50 ~$\mu$m and 100 $\mu$m.}
\end{figure}

Before exploring diffusiophoresis under a large pH gradient, we first set up experiments with a small pH gradient using \ce{HCl} and \ce{NaOH} solutions (separately). Details of all the experiments are described in Appendix A. The diffusiophoretic mobility ($\Gamma_p$) set by a binary electrolyte can be calculated in the limit of negligible double layer thickness as \cite{prieve1984jfm}
\begin{equation}
\Gamma_p = \frac{\epsilon}{\mu}\frac{k_B T}{z e}\left[\beta \zeta_p - \frac{2 k_B T}{z e}\ln \left(1-\tanh^2 \frac{z e \zeta_p}{4 k_B T}\right)\right]~,
\label{MobilityEquation}
\end{equation}
where $\epsilon$, $\mu$, $k_B$ $T$, $z$, and $e$ are, respectively, the electrical permittivity, fluid viscosity, Boltzmann constant, absolute temperature, valence ($z=1$ for \ce{HCl} and \ce{NaOH}), and the charge of an electron. $\beta$ is the diffusivity difference factor, defined as $\displaystyle{\beta=\frac{D_+ - D_-}{D_+ + D_-}}$, where $D_+$ and $D_-$ are, respectively, the diffusion coefficients of the positive and negative ions. The Debye length is defined as $\displaystyle{\lambda_D=\sqrt{\frac{\epsilon k_B T}{2 e^2 c}}}$, where $c$ is the ionic strength. For $c=10$ mM, $\lambda_D \approx 3$ nm, which is negligible compared to the size of particles used in the study. Therefore, equation (\ref{MobilityEquation}) can be used to estimate the diffusiophoretic mobility of micron-sized polystyrene particles. The parameters used for the calculations are organized in Table \ref{table1}. The mobility, equation (\ref{MobilityEquation}), is plotted versus zeta potential $\zeta_p$ in Figure \ref{fig1}(a). Within the range of $\zeta_p$ plotted, $\Gamma_p$ for separate solutions of HCl and NaOH changes sign when the sign of $\zeta_p$ changes. In \ce{HCl}, positively charged particles move up the concentration gradient, whereas in \ce{NaOH} solution, negatively charged particles move up the concentration gradient. 

We observe the same trend in the compaction experiments in a dead-end pore geometry with amine-modified polystyrene (a-PS) and polystyrene (PS) particles (Figure \ref{fig1}(b,c)). Dead-end pores with width, height, and length, respectively, $w=100~\mu$m, $h=50~\mu$m, and $\ell=1$ mm are initially filled with a particle suspension (initial electrolyte concentration in the pore $c_p$ = 10 mM). An air bubble is used as a spacer, then an aqueous solution without any particles is flowed in the main channel and connected with the liquid in the pores (channel electrolyte concentration $c_c$= 1 mM). We use $c_p$ for the initial concentration of chemical species in the pore, and $c_c$ for the concentration in the main channel throughout the paper. Under a concentration gradient of \ce{HCl}, we observe that the amine-modified polystyrene (a-PS, diameter = 1 $\mu$m) particles compact toward the dead-end. In the \ce{NaOH} concentration gradient, also as expected, the polystyrene (PS, diameter = 1 $\mu$m) particles move toward the dead-end. In one situation where a-PS particles are initially suspended in a 10 mM NaOH solution (initial pH in the pore is pH$_p =12$; Figure \ref{fig1}(d)), the time-varying concentration gradient set by the 1 mM NaOH in the main channel (pH$_c = 11$) made the particles move toward the dead-end, which indicates that a-PS particles behave like negatively charged polystyrene particles in such a configuration (see Video S1).

The experimental observations in Figure \ref{fig1}  naturally led us to measure the zeta potential of the a-PS particles. Zeta potentiometry (Anton Paar Litesizer 500 \cite{patent}) is done using 10 mM \ce{NaCl} as a background electrolyte, and the pH is varied by adding \ce{HCl} or \ce{NaOH}. For pH 2 and 12, \ce{NaCl} is not added so that the ionic strength of all suspensions is $\approx 10$ mM. For two batches of the a-PS particles (Sigma Aldrich L9654; MKCF6014 and MKCK7640), we obtain almost identical trends of zeta potential for different pH (Figure \ref{fig2}(a)). We note that there is a sharp decrease and a sign change in the zeta potential of a-PS particles between pH=11 and pH=12 (pI $\approx 11.6$). In contrast, polystyrene particles (Invitrogen F13082) stay negatively charged within a wide range of pH, with some variations in the magnitudes of the potential. Diffusiophoresis of the negatively charged polstyrene particles in a pH gradient has been studied systematically by Shi et al. \cite{squires2016}, so we focus on the experimental measurements made with a-PS particles (with a pI) throughout the rest of the article.

\section{pH dependent zeta potential of a-PS particles}
\label{SectionIII}

\begin{figure}[t!]
\centering
\includegraphics[width=0.96\textwidth]{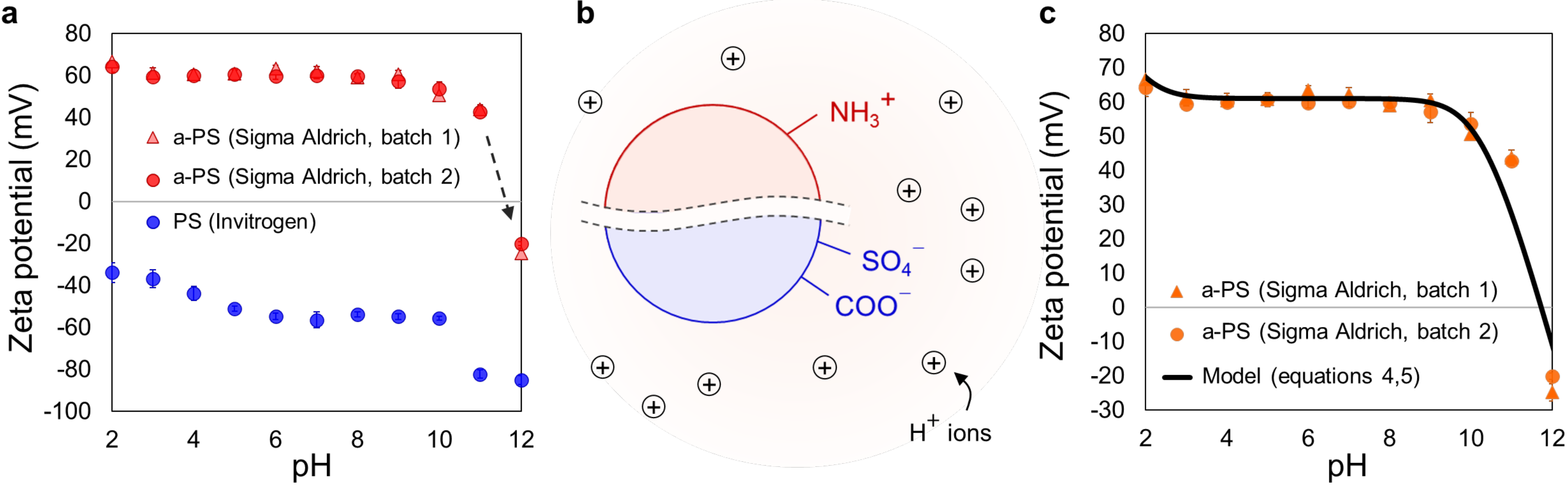}
\caption{\label{fig2} Zeta potential of PS and a-PS particles. (a) Zeta potential measurements (Anton Paar Litesizer 500) for PS and a-PS particles plotted versus pH. We observe a sharp change in the zeta potential (with a sign change) of a-PS particles between pH=11 and pH=12 (pI $\approx$ 11.6). (b) Schematic of acidic and basic surface functional groups on a polystyrene particle. Depending on the ionization of the functional groups, we obtain different equilibrium surface potential values. (c) A charge regulation model (equations 4,5) yields the zeta potential of a-PS particles as a function of pH, and fit the measured data.  }
\end{figure}

In order to obtain the form of the zeta potential of a-PS as a function of \ce{H+} concentration in the bulk, we formulate a charge regulation model that considers acidic and basic surface groups (Figure \ref{fig2}(b)) \cite{healy1978ionizable,healy1985,healy1992,saville1992}. Acidic functional groups such as sulfate or carboxylate can be present on commercial microspheres as part of the polymerization process or as a consequence of additional modification steps \cite{sigma, thermofisher}. Such functional groups (\ce{HA}) follow the typical reaction $\ce{HA} \leftrightharpoons \ce{H+} + \ce{A-}$, so that \ce{A-} contributes to the negative surface charge. Considering the reaction equation and including the Boltzmann distribution of ions near the surface, we can count the number of charge-contributing acidic surface groups and obtain the charge density $q_A$ as (see Appendix B for details)
\begin{equation}
q_{A}=-e\ce{[A^-]} = \frac{-e \ce{n_A}}{1+10^{\ce{pK_A}-\ce{pH}} \exp\left(-\frac{\epsilon \zeta_p}{k_B T}\right)}~.
\end{equation}
Here, \ce{n_A} and \ce{K_A} are, respectively, the total number density (per area) of the acidic surface groups ($\ce{n_A}=\ce{[HA]}+\ce{[A^-]}$) and the acid dissociation constant, where $\ce{pK_A} =-\log_{10}\ce{K_A}$. 

Similarly, basic surface groups can be counted. Basic functional groups (e.g., \ce{NH2}) follow the reaction $\ce{BH^+} \leftrightharpoons \ce{B}+\ce{H+}$, and \ce{[BH^+]} contributes to the positive surface charge. By defining $\ce{K_B}$ as the acid dissociation constant of the conjugate acid $\ce{BH+}$, with $\ce{pK_B} =-\log_{10}\ce{K_B}$, and $\ce{n_B}=[\ce{BH+}] +[\ce{B}]$ as the total number density of the basic functional groups, we obtain the positive charge density $q_B$ as
\begin{equation}
q_{B}=e[\ce{BH+}] = \frac{e \ce{n_B} 10^{\ce{pK_B}-\ce{pH}} \exp \left(-\frac{e \zeta_p}{k_B T}\right)}{1+ 10^{\ce{pK_B}-\ce{pH}} \exp \left(-\frac{e \zeta_p}{k_B T}\right)}~.
\end{equation}

The a-PS particles used in this study \cite{sigma} show varying zeta potential near pH=3 and pH=12, which means that the carboxylate surface group is not likely to be present on the particle surface as its \ce{pK_A}=5. Therefore, we assume that the surface of amine-modified polystyrene has sulfate and amine functional groups. The manufacturers do not share details about microsphere fabrication, but our assumption appears reasonable according to the explanations provided in technical notes distributed by the companies \cite{sigma, thermofisher}. Either sulfate or carboxylate functional groups can be used during the polymerization of styrene, and then the amine modification step is applied to the polystyrene particles \cite{thermofisher}. The net surface charge density $q=q_A + q_B$ thus can be described as 
\begin{equation}
q=-e[\ce{SO4-}]+e[\ce{NH3+}] = \frac{-e \ce{n_A}}{1+10^{\ce{pK_A}-\ce{pH}} \exp\left(-\frac{\epsilon \zeta_p}{k_B T}\right)} + \frac{e \ce{n_B} 10^{\ce{pK_B}-\ce{pH}} \exp \left(-\frac{e \zeta_p}{k_B T}\right)}{1+ 10^{\ce{pK_B}-\ce{pH}} \exp \left(-\frac{e \zeta_p}{k_B T}\right)}~\label{eqncr}.
\end{equation}
For sulfate groups \ce{pK_A=2} and \ce{pK_B} of \ce{NH3+} is not known for this specific product. A technical document from Sigma Aldrich mentions that the surface coverage of the functional groups is estimated as 30-300 \AA$^2$ per charge group \cite{sigma}.

The surface charge density is balanced with the equilibrium potential in the electrical double layer (EDL). In our zeta potential measurements, the thickness of the EDL is determined by the ionic strength of 10 mM NaCl (for pH=2 it is 10 mM HCl, and for pH=12 it is 10 mM NaOH). Therefore, we can use the Gouy-Chapmann formulation for a binary system to relate the zeta potential $\zeta_p$ and the surface charge density $q$,
\begin{equation}
q = 4 c e\lambda_D  \sinh \left(\frac{e \zeta_p}{2 k_B T}\right)~\label{eqngc}.
\end{equation}
The Debye length is defined as $\displaystyle{\lambda_D = \sqrt{\frac{\epsilon k_B T}{2 e^2 c}}}$, where $c$ is the 10 mM ionic strength (see Table \ref{table1} for the parameters used for calculations). We obtain the zeta potential as a function of pH by equating the equations (\ref{eqncr}) and (\ref{eqngc}) with fitting parameters $e$\ce{n_A} $= 0.0402$ C/m$^2$, $e$\ce{n_B} $=0.0576$ C/m$^2$ and \ce{pK_B} $=12.1$. A least squares fit is used with $e$\ce{n_A} $=\pm0.0001$ C/m$^2$, $e$\ce{n_B} $=\pm 0.0001$ C/m$^2$, and \ce{pK_B} $=\pm0.05$, with a condition $\zeta_p(\ce{pH}=12)<-10$ mV. The solution is compared with the measured data in Figure \ref{fig2}(c). The number density $\ce{n_A}+\ce{n_B} = 6.11 \times 10^{17}$ per m$^2$ corresponds to 164  \AA$^2$ per charge group, and is consistent with the values given in the manufacturer's technical note \cite{sigma}. We will use this solution in the calculations for diffusiophoresis in later sections.

\section{Diffusiophoresis of a-PS particles in a dead-end pore geometry}
\label{SectionIV}
\subsection{Experiments}

\begin{figure}[b!]
\centering
\includegraphics[width=0.93\textwidth]{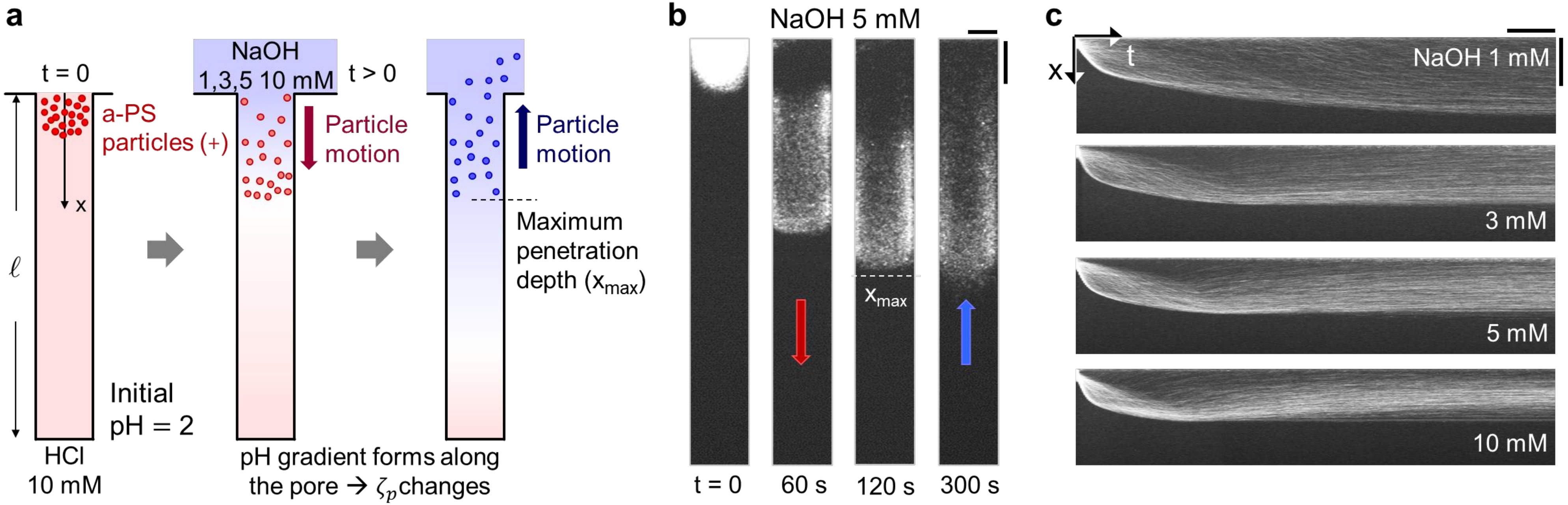}
\caption{\label{fig3} Dead-end pore experiments with different combinations of initial pH in the pores (pH$_p$) and maintained pH outside the pore (pH$_c$). (a) Schematics describing the experiments. The pore is initially filled with 2 mM \ce{HCl} (pH$_p=2$). Then a patch of a-PS particles (pI $\approx 11.6$) are introduced near the inlet. We indicate the initial positive charge of the particles as (+). The particles undergo diffusiophoresis along the concentration gradients created by 1, 3, 5, and 10 mM NaOH solutions, which flow in the main channel. (b) Time sequence images of a typical experiment. The a-PS particles are initially positively charged in the \ce{HCl} solution, so migrate into the pore by diffusiophoresis. Then the front of the particle patch reaches a maximum penetration depth ($x_{\text{max}}$), and changes direction of motion toward the pore inlet. When the particles are moving toward $x_{\text{max}}$, we observe that the front is nearly flat, showing that the influence of diffusioosmosis is small. Horizontal and vertical scale bars are, respectively, 50 $\mu$m and 100 $\mu$m. (c) The particle  trajectories along the centerline ($x(t)$) are visualized for all four chemical conditions. Horizontal and vertical scale bars are, respectively, 60 seconds and 500 $\mu$m. } 
\end{figure}

By measuring and calculating the zeta potential of a-PS particles as a function of pH, we confirm, for the Sigma Aldrich a-PS particles, the existence of the isoelectric point (pI) near pH=11 and 12 (pI $\approx 11.6$). Next, using a dead-end pore geometry, we design a set of experiments to systematically study the diffusiophoresis of these particles. In the same dead-end pore used in earlier experiments ($w=100~\mu$m, $h=50~\mu$m and $\ell=1$ mm), we create concentration gradients of ions by initially filling the pore with 10 mM \ce{HCl} solution, then introducing  1, 3, 5 and 10 mM \ce{NaOH} solutions (respectively) in the main channel. In this way, the initial pH inside the pore is fixed at pH$_p$=2, and the pH in the main channel is set in each experiment at pH$_c = 11,~11.5,~11.7$, and 12, respectively. Particles are initially localized near the inlet so that the initial penetration depth of the particle patch is $x\approx w$. Details of the experimental setup is described in the Appendix A. Graphical explanations for the setup can be found in Alessio et al. \cite{alessio2022jfm}. 

The liquid stream that creates the particle patch is followed by an air bubble (a spacer), then the NaOH solution. Once the NaOH and HCl solutions come in contact, diffusion of ions and diffusiophoresis of particles occur along the pores (Figure \ref{fig3}(a)). Except for the case with 1 mM NaOH solution, we observe that the penetration of the particles reaches a maximum distance. At early times, a particle patch propagates toward the dead-end of the pores by first spreading and then focusing near the front. Until the patch reaches its maximum penetration depth, the front remains flat, meaning that the influence of wall diffusioosmosis is negligible at the locations and times of the front. After the maximum penetration, the direction of particle motion changes, and the influence of diffusioosmosis is non-negligible. At this later time, ions have diffused enough so that the contribution of wall diffusioosmosis on the reverse path of the particles is different from that on the forward path where the front was translating toward fresh 10 mM HCl solution (Figure \ref{fig3}(b) and Video S2). The influence of diffusioosmosis changes because the PDMS walls also have a pH-dependent zeta potential \cite{kirby2004-2} (see Appendix \ref{SectionAC} for details).

We analyze the early translation of the particle front using one-dimensional (1D) multi-ion diffusiophoresis calculations. As the main control parameter is the varying zeta potential of particles along a pH gradient, we focus on the regime where diffusioosmosis is negligible. From the experimental images, we obtain a kymograph (ImageJ) to plot the centerline data versus time. In Figure \ref{fig3}(c), the time evolution of the particle patch along the centerline of the pore is visualized for all four experimental conditions. For the case with NaOH concentration $c_c=1$ mM, there is no stopping of particles or maximum penetration depth. Particles maintain their positive surface potential throughout the experiment, and continue to migrate into the pore within the observation time (600 s). When $c_c=3$ and $5$ mM, we observe that the front reaches a maximum penetration depth ($x_{\text{max}}$). In both cases, the front position does not change after maximum penetration, but the particle patch widens over time, showing that the particles closer to the inlet move back toward the main channel. The change in the late time particle distribution is due to the non-negligible wall diffusioosmosis. Finally, when $c_c=10$ mM, we observe that the front reaches $x_{\text{max}}$, and reverses its direction of motion, while experiencing the widening of patch due to the wall diffusioosmosis. Next, we consider the diffusion-reaction of multiple ions in the 1D system.

\subsection{Mathematical model for multi-ion diffusiophoresis}

In a 1D pore, the ion transport for concentration $c_i (x,t)$ is described with the Nernst-Planck equation (the $x$ axis is defined from the pore inlet) \cite{gupta2019prf,alessio2021prf}
\begin{equation}
    \frac{\partial c_i}{\partial t}=D_i \frac{\partial^2 c_i}{\partial x^2}+\frac{D_i z_i e}{k_B T}\left(\frac{\partial c_i}{\partial x}\frac{\partial \psi}{\partial x} + c_i \frac{\partial^2 \psi}{\partial x^2}\right) + \mathcal{R}~,
\end{equation}
where the subscript $i$ is for different ions. Here,  $z_i$, $\psi$, and $\mathcal{R}$ are, respectively, the valence of the $i^{\text{th}}$ ion, electric potential, and the chemical reaction term. Rewriting the equation for each ion in a solution containing NaOH and HCl, we obtain
\begin{subequations}
\begin{align}
    \frac{\partial c_{\scaleto{\ce{Na}}{4pt}}}{\partial t} &= D_{\scaleto{\ce{Na}}{4pt}} \frac{\partial^2 c_{\scaleto{\ce{Na}}{4pt}}}{\partial x^2} + \frac{D_{\scaleto{\ce{Na}}{4pt}}e}{k_B T} \left( \frac{\partial c_{\scaleto{\ce{Na}}{4pt}}}{\partial x} \frac{\partial \psi}{\partial  x} + c_{\scaleto{\ce{Na}}{4pt}}\frac{\partial^2 \psi}{\partial x^2}\right)\\
    \frac{\partial c_{\scaleto{\ce{Cl}}{4pt}}}{\partial t} &= D_{\scaleto{\ce{Cl}}{4pt}}\frac{\partial^2 c_{\scaleto{\ce{Cl}}{4pt}}}{\partial x^2} - \frac{D_{\scaleto{\ce{Cl}}{4pt}} e}{k_B T} \left( \frac{\partial c_{\scaleto{\ce{Cl}}{4pt}}}{\partial x} \frac{\partial \psi}{\partial x} + c_{\scaleto{\ce{Cl}}{4pt}}\frac{\partial^2 \psi}{\partial x^2} \right) \\
    \frac{\partial c_{\scaleto{\ce{H}}{4pt}}}{\partial t} &= D_{\scaleto{\ce{H}}{4pt}}\frac{\partial^2 c_{\scaleto{\ce{H}}{4pt}}}{\partial x^2} + \frac{D_{\scaleto{\ce{H}}{4pt}}e}{k_B T} \left( \frac{\partial c_{\scaleto{\ce{H}}{4pt}}}{\partial x} \frac{\partial \psi}{\partial  x} + c_{\scaleto{\ce{H}}{4pt}}\frac{\partial^2 \psi}{\partial x^2}\right) - k_r (c_{\scaleto{\ce{H}}{4pt}} c_{\scaleto{\ce{OH}}{4pt}} - K_w)\\
    \frac{\partial c_{\scaleto{\ce{OH}}{4pt}}}{\partial t} &= D_{\scaleto{\ce{OH}}{4pt}}\frac{\partial^2 c_{\scaleto{\ce{OH}}{4pt}}}{\partial x^2} - \frac{D_{\scaleto{\ce{OH}}{4pt}}e}{k_B T} \left( \frac{\partial c_{\scaleto{\ce{OH}}{4pt}}}{\partial x} \frac{\partial \psi}{\partial  x} + c_{\scaleto{\ce{OH}}{4pt}}\frac{\partial^2 \psi}{\partial x^2}\right) - k_r (c_{\scaleto{\ce{H}}{4pt}} c_{\scaleto{\ce{OH}}{4pt}} - K_w)~,
\end{align}
\end{subequations}
where $D_{\scaleto{\ce{Na}}{4pt}}$, $D_{\scaleto{\ce{Cl}}{4pt}}$, $D_{\scaleto{\ce{H}}{4pt}}$, and $D_{\scaleto{\ce{OH}}{4pt}}$ and $c_{\scaleto{\ce{Na}}{4pt}}$, $c_{\scaleto{\ce{Cl}}{4pt}}$, $c_{\scaleto{\ce{H}}{4pt}}$, and $c_{\scaleto{\ce{OH}}{4pt}}$ are, respectively, the diffusivity and concentration of \ce{Na+}, \ce{Cl-}, \ce{H+}, and \ce{OH-}. Also, $k_r$ and $K_w$ are, respectively, the backward reaction constant and the equilibrium constant of the autoionization reaction $\ce{H2O}(\ell) \leftrightharpoons \ce{H+}+\ce{OH-}$.

The initial and boundary conditions are 
\begin{equation}
    c_{i}(x,0) = c_{ip}~,~~c_{i}(0,t) = c_{ic}~,~~\frac{\partial c_i}{\partial x}\big|_{x=\ell} = 0    ,
\end{equation}
where the initial values for the pore $c_{ip}$ and channel $c_{ic}$ are varied to match the experimental conditions; typically, the initial concentration of \ce{HCl} in the pores is 10 mM and the concentrations of \ce{NaOH} in the main channel are 1, 3, 5, and 10 mM.   

\begingroup
\renewcommand{\arraystretch}{1.11} 
\begin{table}[b!]
\centering
\begin{tabular}{ P{3cm} P{4cm} m{10cm} }
 Variable & Quantity & Description \\ 
\hline
 $\beta_{\ce{HCl}}$ & 0.642 & Diffusivity difference factor of \ce{HCl}   \\ 
 $\beta_{\ce{NaOH}}$ & -0.596 & Diffusivity difference factor of \ce{NaOH}   \\ 
 $\beta_{\ce{NaCl}}$ & -0.207 & Diffusivity difference factor of \ce{NaCl}   \\ 
 $\beta_{\ce{KCl}}$ & -0.019 & Diffusivity difference factor of \ce{KCl}    \\
 $\epsilon$ & 7.0 $\times 10^{-10}$ F/m& Dielectric permittivity for dilute  electrolyte solutions* \\
 $\mu$	    & $10^{-3}$ Pa$\cdot$s & Dynamic viscosity of aqueous solutions (HCl, NaOH, NaCl, KCl)**\\
 $T$ & 298 K& Absolute temperature \\
 $\ell$  & 1 mm & Length of the dead-end pore \\
 $D_{\scaleto{\ce{H}}{4pt}}$ & $9.311 \times 10^{-9}$ m$^2$/s & Diffusion coefficient of \ce{H+}\\
 $D_{\scaleto{\ce{OH}}{4pt}}$ & $5.273 \times 10^{-9}$ m$^2$/s & Diffusion coefficient of \ce{OH-}\\
 $D_{\scaleto{\ce{Na}}{4pt}}$ & $1.334 \times 10^{-9}$ m$^2$/s & Diffusion coefficient of \ce{Na+}\\
 $D_{\scaleto{\ce{Cl}}{4pt}}$ & $2.032 \times 10^{-9}$ m$^2$/s & Diffusion coefficient of \ce{Cl-}\\
 $k_r$ & $1.4 \times 10^{-11}$ M$^{-1}$s$^{-1}$ & Rate constant of water reaction \cite{waterrxn} \\
 $K_w$ & 10$^{-14}$ M$^2$ & Equilibrium constant of water reaction \cite{waterrxn}\\
\hline
\end{tabular}
\caption{Parameters used in the calculations. The values for diffusivities and $\beta$ are obtained from Velegol et al. \cite{velegol2016review}. *Literature references \cite{epswater,epshcl,epsnaoh,epsnacl,epskcl} suggest that for dilute solutions (10-20 mM \ce{HCl}, \ce{NaOH}, \ce{NaCl}, and \ce{KCl}) relative permittivities are 78-80, and thus we use a representative value $\epsilon = 7.0 \times 10^{-10}$ F/m for all calculations. **Also a representative value is chosen for all calculations. }
\label{table1}
\end{table}
\endgroup

Electroneutrality is maintained in the pore throughout the experiments, and thus
\begin{equation}
    \sum_{i}{z_i c_i} = 0~~~ \Rightarrow ~~~ c_{\scaleto{\ce{Na}}{4pt}}-c_{\scaleto{\ce{Cl}}{4pt}}+c_{\scaleto{\ce{H}}{4pt}}-c_{\scaleto{\ce{OH}}{4pt}}=0~.
\end{equation}
Also, we have the zero current condition $\sum_{i}{z_i {j_i}}=0$, where $\displaystyle{{j_i} =- D_i\left(\frac{\partial c_i}{\partial x} + \frac{z_i e c_i}{k_B T}\frac{\partial \psi}{\partial x}\right)}$, so we obtain 
\begin{equation}
    \frac{\partial \psi}{\partial x}=-\frac{k_B T}{e} \frac{\sum_{i}{D_i z_i \frac{\partial c_i}{\partial x}}}{\sum_{i}{D_i z_i^2 c_i}} = -\frac{k_B T}{e}\left( \frac{D_{\scaleto{\ce{Na}}{4pt}}\frac{\partial c_{\scaleto{\ce{Na}}{3pt}}}{\partial x}-D_{\scaleto{\ce{Cl}}{4pt}}\frac{\partial c_{\scaleto{\ce{Cl}}{3pt}}}{\partial x}+D_{\scaleto{\ce{H}}{4pt}}\frac{\partial c_{\scaleto{\ce{H}}{3pt}}}{\partial x}-D_{\scaleto{\ce{OH}}{4pt}}\frac{\partial c_{\scaleto{\ce{OH}}{3pt}}}{\partial x}}{D_{\scaleto{\ce{Na}}{4pt}}c_{\scaleto{\ce{Na}}{4pt}}+D_{\scaleto{\ce{Cl}}{4pt}}c_{\scaleto{\ce{Cl}}{4pt}}+D_{\scaleto{\ce{H}}{4pt}}c_{\scaleto{\ce{H}}{4pt}}+D_{\scaleto{\ce{OH}}{4pt}}c_{\scaleto{\ce{OH}}{4pt}}}\right)~.
\end{equation}
The diffusiophoretic velocity is \cite{gupta2019prf}
\begin{align}
    u_p &= \frac{\epsilon}{\mu} \left(\frac{k_B T}{e} \right) \frac{\sum_{i}{D_i z_i \frac{\partial c_i}{\partial x}}}{\sum_{i}{D_i z_i^2 c_i}} \zeta_p + \frac{\epsilon}{8 \mu} \frac{\sum_{i}{z_i^2 \frac{\partial c_i}{\partial x}}}{\sum_{i}{z_i^2 c_i}} \zeta_p^2\\ 
    &= \frac{\epsilon}{\mu} \left( \frac{k_B T}{e}\right) \frac{D_{\scaleto{\ce{Na}}{4pt}}\frac{\partial c_{\scaleto{\ce{Na}}{3pt}}}{\partial x}-D_{\scaleto{\ce{Cl}}{4pt}}\frac{\partial c_{\scaleto{\ce{Cl}}{3pt}}}{\partial x}+D_{\scaleto{\ce{H}}{4pt}}\frac{\partial c_{\scaleto{\ce{H}}{3pt}}}{\partial x}-D_{\scaleto{\ce{OH}}{4pt}}\frac{\partial c_{\scaleto{\ce{OH}}{3pt}}}{\partial x}}{D_{\scaleto{\ce{Na}}{4pt}}c_{\scaleto{\ce{Na}}{4pt}}+D_{\scaleto{\ce{Cl}}{4pt}}c_{\scaleto{\ce{Cl}}{4pt}}+D_{\scaleto{\ce{H}}{4pt}}c_{\scaleto{\ce{H}}{4pt}}+D_{\scaleto{\ce{OH}}{4pt}}c_{\scaleto{\ce{OH}}{4pt}}} \zeta_p + \frac{\epsilon}{ 8\mu} \frac{\frac{\partial c_{\scaleto{\ce{Na}}{3pt}}}{\partial x}+\frac{\partial c_{\scaleto{\ce{Cl}}{3pt}}}{\partial x}+\frac{\partial c_{\scaleto{\ce{H}}{3pt}}}{\partial x}+\frac{\partial c_{\scaleto{\ce{OH}}{3pt}}}{\partial x}} {c_{\scaleto{\ce{Na}}{4pt}}+c_{\scaleto{\ce{Cl}}{4pt}}+c_{\scaleto{\ce{H}}{4pt}}+c_{\scaleto{\ce{OH}}{4pt}}} \zeta_p^2~.
\end{align}

The equations are nondimensionalized by defining 
\begin{equation}
    \bar{c}_i = \frac{c_i}{c^*}~,~~\bar{D}_i = {\frac{D_i}{D_{\scaleto{\text{H}}{4pt}}}}~,~~X=\frac{x}{\ell}~,~~ \Psi=\frac{\psi e}{k_B T}~,~~\bar{\zeta}_p = \frac{\zeta_p e}{k_B T}~,~~\tau={\frac{t}{\ell^2/D_{\scaleto{\text{H}}{4pt}}}}~,~~\bar{U}_p = {\frac{u_p}{D_{\scaleto{\text{H}}{4pt}}/\ell}}.
\end{equation}
The characteristic concentration $c^*$ is set as $c^*=1$ M for convenience.

The nondimensional form of (7) gives
\begin{subequations}
\begin{align}
    &\frac{\partial \bar{c}_{\scaleto{\ce{Na}}{4pt}}}{\partial \tau} = \bar{D}_{\scaleto{\ce{Na}}{4pt}} \left(\frac{\partial^2 \bar{c}_{\scaleto{\ce{Na}}{4pt}}}{\partial X^2} + \frac{\partial \bar{c}_{\scaleto{\ce{Na}}{4pt}}}{\partial X} \frac{\partial \Psi}{\partial X} + \bar{c}_{\scaleto{\ce{Na}}{4pt}}\frac{\partial^2 \Psi}{\partial X^2} \right) \label{eqcna}\\
    &\frac{\partial\bar{c}_{\scaleto{\ce{Cl}}{4pt}}}{\partial \tau} = \bar{D}_{\scaleto{\ce{Cl}}{4pt}} \left( \frac{\partial^2 \bar{c}_{\scaleto{\ce{Cl}}{4pt}}}{\partial X^2} - \frac{\partial \bar{c}_{\scaleto{\ce{Cl}}{4pt}}}{\partial X} \frac{\partial \Psi}{\partial X} - \bar{c}_{\scaleto{\ce{Cl}}{4pt}} \frac{\partial^2 \Psi}{\partial X^2} \right) \label{eqccl}\\
    &\frac{\partial \bar{c}_{\scaleto{\ce{H}}{4pt}}}{\partial \tau} = \frac{\partial^2 \bar{c}_{\scaleto{\ce{H}}{4pt}}}{\partial X^2} + \frac{\partial \bar{c}_{\scaleto{\ce{H}}{4pt}}}{\partial X}\frac{\partial \Psi}{\partial X} + \bar{c}_{\scaleto{\ce{H}}{4pt}} \frac{\partial^2 \Psi}{\partial X^2} - K_r (\bar{c}_{\scaleto{\ce{H}}{4pt}} \bar{c}_{\scaleto{\ce{OH}}{4pt}} - \bar{K}_w ) \label{eqch}\\
    &\frac{\partial \bar{c}_{\scaleto{\ce{OH}}{4pt}}}{\partial \tau} =\bar{D}_{\scaleto{\ce{OH}}{4pt}} \left( \frac{\partial^2 \bar{c}_{\scaleto{\ce{OH}}{4pt}}}{\partial X^2} - \frac{\partial \bar{c}_{\scaleto{\ce{OH}}{4pt}}}{\partial X} \frac{\partial \Psi}{\partial X} - \bar{c}_{\scaleto{\ce{OH}}{4pt}} \frac{\partial^2 \Psi}{\partial X^2} \right) - K_r(\bar{c}_{\scaleto{\ce{H}}{4pt}} \bar{c}_{\scaleto{\ce{OH}}{4pt}} - \bar{K}_w ) \label{eqcoh}\\
    &\bar{c}_{\scaleto{\ce{Na}}{4pt}}-\bar{c}_{\scaleto{\ce{Cl}}{4pt}}+\bar{c}_{\scaleto{\ce{H}}{4pt}}-\bar{c}_{\scaleto{\ce{OH}}{4pt}}=0 \label{eqneutral}\\
    &\frac{\partial \Psi}{\partial X} = -\frac{\bar{D}_{\scaleto{\ce{Na}}{4pt}}\frac{\partial \bar{c}_{\scaleto{\ce{Na}}{3pt}}}{\partial X}-\bar{D}_{\scaleto{\ce{Cl}}{4pt}}\frac{\partial \bar{c}_{\scaleto{\ce{Cl}}{3pt}}}{\partial X}+\frac{\partial \bar{c}_{\scaleto{\ce{H}}{3pt}}}{\partial X}-\bar{D}_{\scaleto{\ce{OH}}{4pt}}\frac{\partial \bar{c}_{\scaleto{\ce{OH}}{3pt}}}{\partial X}}{\bar{D}_{\scaleto{\ce{Na}}{4pt}}\bar{c}_{\scaleto{\ce{Na}}{4pt}}+\bar{D}_{\scaleto{\ce{Cl}}{4pt}}\bar{c}_{\scaleto{\ce{Cl}}{4pt}}+\bar{c}_{\scaleto{\ce{H}}{4pt}}+\bar{D}_{\scaleto{\ce{OH}}{4pt}}\bar{c}_{\scaleto{\ce{OH}}{4pt}}} \label{eqpotential}\\
    &\bar{U}_p = \frac{\epsilon}{\mu}\left(\frac{k_B T}{e} \right)^2 \frac{1}{D_{\ce{H}}} \left[ \frac{\bar{D}_{\scaleto{\ce{Na}}{4pt}}\frac{\partial \bar{c}_{\scaleto{\ce{Na}}{3pt}}}{\partial X}-\bar{D}_{\scaleto{\ce{Cl}}{4pt}}\frac{\partial \bar{c}_{\scaleto{\ce{Cl}}{3pt}}}{\partial X}+\frac{\partial \bar{c}_{\scaleto{\ce{H}}{3pt}}}{\partial X}-\bar{D}_{\scaleto{\ce{OH}}{4pt}}\frac{\partial \bar{c}_{\scaleto{\ce{OH}}{3pt}}}{\partial X}}{\bar{D}_{\scaleto{\ce{Na}}{4pt}}\bar{c}_{\scaleto{\ce{Na}}{4pt}}+\bar{D}_{\scaleto{\ce{Cl}}{4pt}}\bar{c}_{\scaleto{\ce{Cl}}{4pt}}+\bar{c}_{\scaleto{\ce{H}}{4pt}}+\bar{D}_{\scaleto{\ce{OH}}{4pt}}\bar{c}_{\scaleto{\ce{OH}}{4pt}}} \bar{\zeta}_p + \frac{1}{8} \frac{\frac{\partial \bar{c}_{\scaleto{\ce{Na}}{3pt}}}{\partial X}+\frac{\partial \bar{c}_{\scaleto{\ce{Cl}}{3pt}}}{\partial X}+\frac{\partial \bar{c}_{\scaleto{\ce{H}}{3pt}}}{\partial X}+\frac{\partial \bar{c}_{\scaleto{\ce{OH}}{3pt}}}{\partial X}}{\bar{c}_{\scaleto{\ce{Na}}{4pt}}+\bar{c}_{\scaleto{\ce{Cl}}{4pt}}+\bar{c}_{\scaleto{\ce{H}}{4pt}}+\bar{c}_{\scaleto{\ce{OH}}{4pt}}} \bar{\zeta}_p^2 \right]~\label{eqvel},
\end{align}
\end{subequations}
where $\displaystyle{K_r = \frac{k_r \ell^2 c^*}{D_{\scaleto{\ce{H}}{4pt}}}}$ and $\displaystyle{\bar{K}_w = \frac{K_w}{{c^*}^2}}$.

\begin{figure}[t!]
\centering
\includegraphics[width=0.76\textwidth]{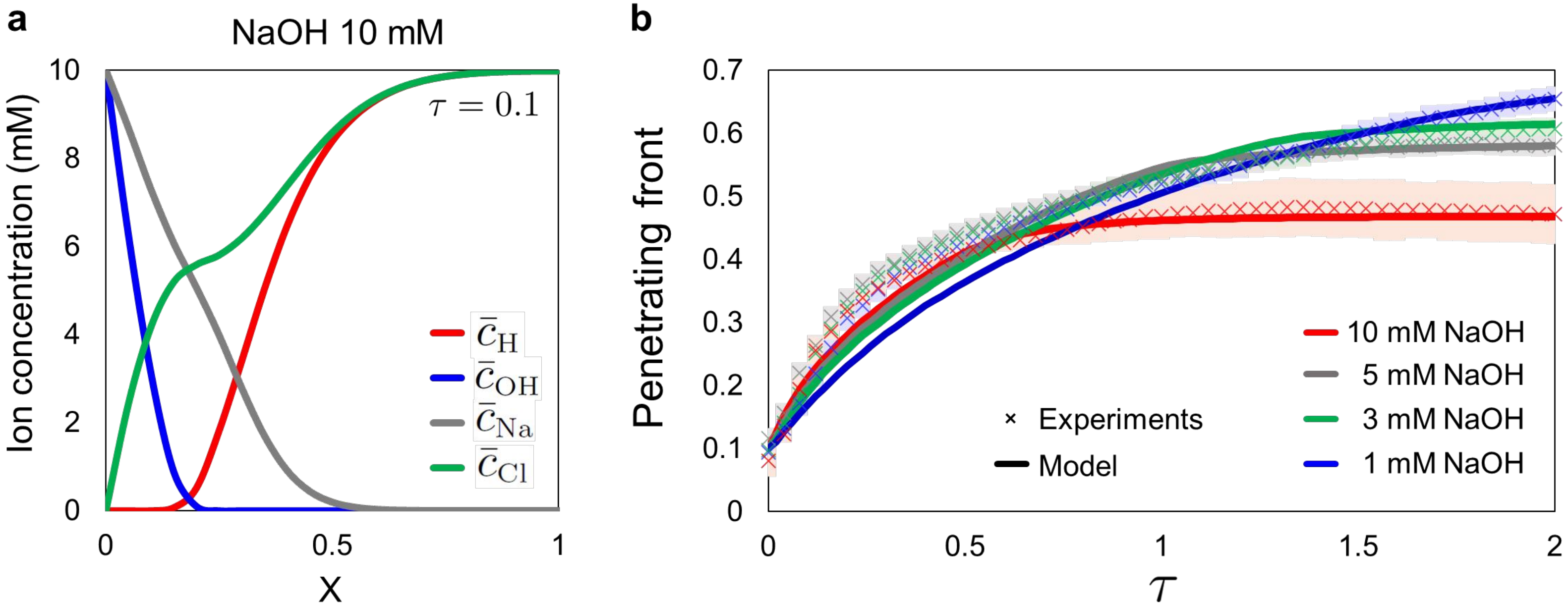}
\caption{\label{fig4} Calculations for diffusion of multiple ions and particle diffusiophoresis in a 1D pore. (a) Calculated ion concentrations for the case with 10 mM NaOH outside the pore and 10 mM HCl (initially) in the pore. As diffusion of ions occurs in the pore, \ce{H+} and \ce{OH-} ions are consumed to produce \ce{H2O} through the (reverse) autoionization reaction. (b) Measured and calculated particle fronts plotted versus nondimensional time ($\tau = t D_{\scaleto{\ce{H}}{4pt}}/\ell^2$) for all four conditions. $\tau=2$ corresponds to $t = 215$ s.}
\end{figure}

The particle front or the 1D trajectory follows 
\begin{equation}
    \frac{\partial X_p}{\partial \tau} = \bar{U}_p (X_p,\tau)~;~~ X_p(\tau=0) = X_0~.\label{eqtraj}
\end{equation}
Equations (\ref{eqcna}-\ref{eqvel}) and (\ref{eqtraj}) are solved numerically with the nondimensional boundary conditions. Calculations for the coupled partial differential equations (PDEs) are done with MATLAB, employing a central difference scheme. For $0<X<1$ and $0<\tau<2$, the spatial and time steps were chosen as, respectively, $\delta X = 0.05$ and $\delta \tau = 10^{-11}$.

The concentrations of the four ions are plotted versus $X$ at $\tau=0.1$ for the case of 10 mM NaOH in Figure \ref{fig4}(a). We note that, as the ions diffuse in the pore, \ce{H+} and \ce{OH-} ions are consumed to produce \ce{H2O}. Different fluxes of ions are combined to contribute to the diffusiophoretic velocity of particles. In experiments, the initial location of the particle front was $x\approx w$, so we calculate the particle trajectory $X_p$ with $X_0=0.1$. The calculated centerline (or 1D particle) trajectories are plotted versus nondimensional time $\tau$ in Figure \ref{fig4}(b) with the experimental data. For $\tau>0.7$ the experimental data and calculations show very good agreement. At early times, due to the merging of two liquid phases (HCl in the pores and NaOH in the main channel), a strong convection that pushes particles into the pore at higher velocity is present near the pore inlet. For 3, 5, 10 mM NaOH, we observe the maximum penetration of the particles within $\tau<2$, and the maximum penetration depth is smallest for 10 mM NaOH (Figures \ref{fig3}(c) and \ref{fig4}(b)). The a-PS particles are always positively charged in 1 mM NaOH solution, and thus the penetration depth keeps increasing without stopping due to the sign change in $\zeta_p$. Of course, after a long time, the ionic fluxes become small and the particle distribution can reach a steady state.

So far, we have identified varying zeta potential of a-PS particles, and obtained a functional form $\zeta_p (\ce{pH})$ by solving the charge regulation model. Such a pH-dependent zeta potential of a-PS particles is visible by the apparent behavior of the particles under a strong pH gradient between pH=2 and pH=12. Particle patch experiments were useful to analyze 1D diffusiophoresis of a-PS particles, but the very first question that motivated the study (Figure \ref{fig1}(d)) is still unanswered. In the compaction experiments of a-PS particles in NaOH solution ($c_p=10$ mM and $c_c=1$ mM), the particle behavior looks similar to that of negatively charged PS. However, the curved compaction boundary suggests that diffusioosmosis along the walls must be included in the analysis \cite{alessio2021prf} in order to correctly interpret the experimental images. In the next section, we present compaction experiments and supporting model calculations for a-PS diffusiophoresis under a pH gradient between pH $=7$ and pH $=12$.

\section{Diffusiophoresis and diffusioosmosis in the presence of a pH gradient}
\label{SectionV}

\begin{figure}[t!]
\centering
\includegraphics[width=\textwidth]{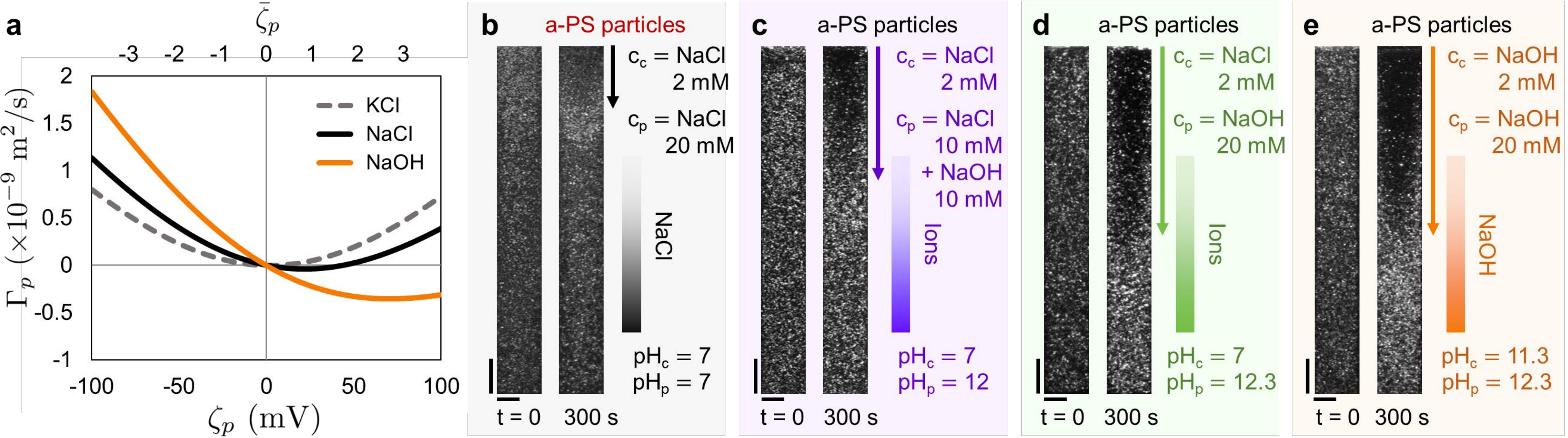}
\caption{\label{fig5} Diffusiophoresis of a-PS particles along a pH gradient set by \ce{NaCl} and \ce{NaOH}. (a) Diffusiophoretic mobilities (equation 1) for \ce{NaCl},\ce{KCl} and \ce{NaOH} plotted versus dimensional ($\zeta_p$) and nondimensional ($\bar{\zeta}_p$) zeta potentials. (b-e) Compaction experiments performed with four different concentration gradients of ions: (b) (I) $c_c=2$ mM NaCl and $c_p=20$ mM NaCl, (c) (II) $c_c=2$ mM NaCl and $c_p=10$ mM NaCl $+ 10$ mM NaOH, (d) (III) $c_c=2$ mM NaCl and $c_p=20$ mM NaOH, and (e) (IV) $c_c=2$ mM NaOH and $c_p$ = 20 mM NaOH. (c-e) When NaOH ($\geq$ 10 mM) is present in the pore, the a-PS particles compacted more toward the dead-end. (b-e) Horizontal and vertical scale bars are, respectively, 50 $\mu$m and 100 $\mu$m.}
\end{figure}

\ce{NaCl} has been one of the most popular salts used in diffusiophoresis studies, due to its convenience and moderate magnitudes of the diffusiophoretic mobilities for commercial PS particles. Often, for negatively charged particles, \ce{NaCl} is considered advantageous over \ce{KCl} due to its higher diffusivity difference factor ($\beta$; see Table \ref{table1}) \cite{velegol2016review}. However, the response of a-PS particles to chemical gradients is more complex when the zeta potential is positive. The diffusiophoretic mobilities (equation (\ref{MobilityEquation})) for \ce{KCl}, \ce{NaCl}, and \ce{NaOH} are plotted versus $\zeta_p$ in Figure \ref{fig5}(a). Within the range of $\zeta_p$ plotted, both \ce{KCl} and \ce{NaCl} appear to be chemiphoresis-dominant (note that chemiphoretic contribution is always positive). The difference between \ce{KCl} and \ce{NaCl} is that, for \ce{KCl}, moderately charged particles ($|{\zeta}_p| \gtrsim 5$ mV) will always move up the concentration gradient, but for \ce{NaCl}, particles with $0 \lesssim {\zeta}_p \lesssim 50$ mV will exhibit  negligible mobility under the chemical gradient. 

For the a-PS particles used in this study,  ${\zeta}_p \approx 60$ mV (or $\bar{\zeta}_p \approx 2.3$) for a wide range of \ce{pH}, which means that \ce{NaCl} diffusiophoresis is not strong when compared to the PS particles (Figure \ref{fig5}(b)). If we alter the initial pH in the dead-end pore by adding \ce{NaOH} ($\geq$ 10 mM; Figure \ref{fig5}(c-e)), we can expect that the particles are initially negatively charged inside the pore and the diffusiophoretic mobility increases. This hypothesis appears to be validated by a set of compaction experiments performed under four different concentration gradients between: (I) $c_c=2$ mM and $c_p=20$ mM \ce{NaCl}, (II) $c_c=2$ mM \ce{NaCl} and $c_p=10$ mM NaCl + 10 mM \ce{NaOH}, (III) $c_c=2$ mM \ce{NaCl} and $c_p=20$ mM \ce{NaOH}, and (IV) $c_c=2$ mM \ce{NaOH} and $c_p=20$ mM \ce{NaOH} (Figure \ref{fig5}(b-e)). However, as we mentioned in Section \ref{SectionIV}, the influence of diffusioosmosis must be included in the analyses of the compaction experiments \cite{alessio2021prf}. For the same set of concentration gradients (I-IV), entrainment experiments are also performed (see Appendix Figure \ref{figA1}).

\begin{figure}[t!]
\centering
\includegraphics[width=0.935\textwidth]{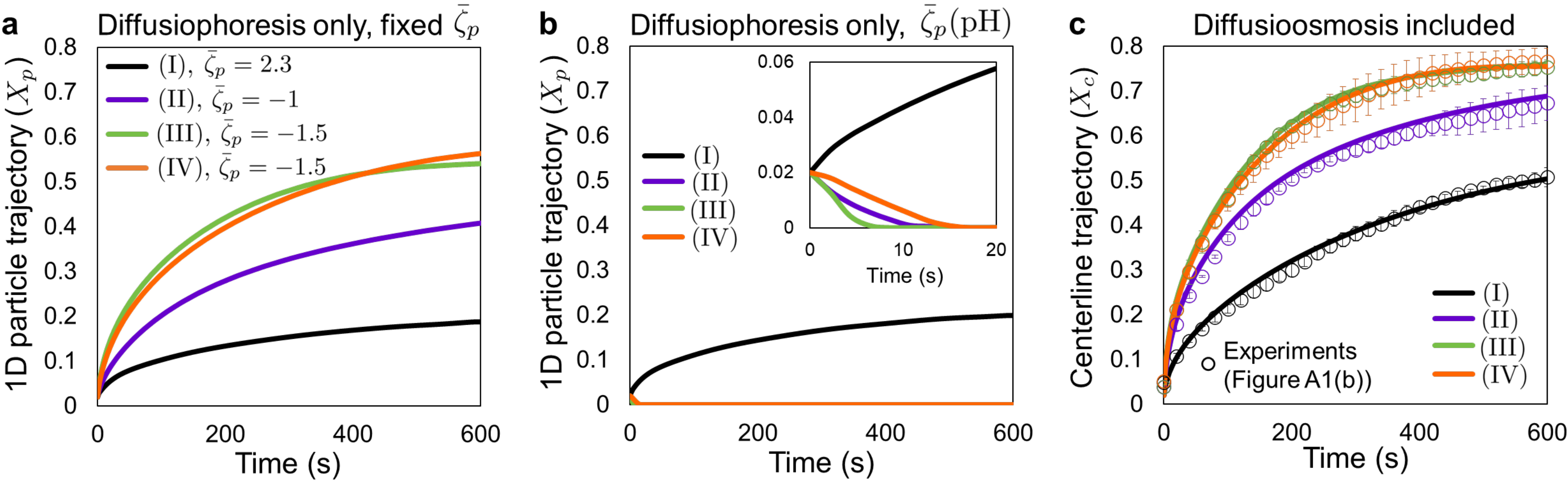}
\caption{\label{fig6} Model calculations for the experimental conditions (I-IV): (I) $c_c=2$ mM and $c_p=20$ mM NaCl, (II) $c_c=2$ mM NaCl and $c_p=10$ mM NaCl + 10 mM NaOH, (III) $c_c=2$ mM NaCl and $c_p=20$ mM NaOH, and (IV) $c_c=2$ mM NaOH and $c_p=20$ mM NaOH.(a) One-dimensional (1D) particle trajectories ($X_p$) are plotted using fixed zeta potentials for four experimental conditions (I-V). Influence of diffusioosmosis is not considered. (b) 1D particle trajectories ($X_p$) considering the varying zeta potential $\zeta_p(\ce{pH})$ (Figure \ref{fig2}(c)) are plotted for four experimental conditions (I-IV). Inset: Early time trajectories. Due to the zeta potential variation, a-PS particles leave the pore by diffusiophoresis when \ce{NaOH} is initially present in the pore (II-IV). (c) Centerline trajectories ($X_c$) including the influence of diffusioosmosis (along the PDMS walls) are plotted for four conditions (I-IV), with the experimental measurements obtained from the entrainment configuration (Figure \ref{figA1}(b)). Only by including the wall diffusioosmosis-driven flow velocity do we obtain the trend for the centerline trajectories that is consistent with the experimental observations. }
\end{figure}

We conducted a set of model calculations to obtain 1D particle  trajectories along the pore centerline ($X_0=0.02$) for scenarios (I-IV), with and without the wall diffusioosmosis (see Appendix \ref{SectionAC} for details). First, we did a test calculation using the initial assumptions for the experimental setup, where the a-PS particles may have a negative surface potential in the presence of \ce{NaOH}, and show increased diffusiophoretic mobilities due to the negative potential. Thus, the test was done for the concentration gradients (I-IV) using fixed zeta potentials (Figure \ref{fig6}(a)), and without including wall diffusioosmosis. Then, we obtain 1D particle trajectories $X_p(\tau)$, as a function of dimensional time, that show the same trend as the experimental images shown in Figure \ref{fig5}(b-e). However, the qualitative agreement in the trend shown in Figure \ref{fig6}(a) does not correctly explain the observations from experiments (I-IV), as experimentally we expect that the zeta potential of a-PS particles vary in the pores. If we use the known zeta potential function ($\zeta_p (\ce{pH})$, Figure \ref{fig2}(c)) to calculate 1D particle trajectories without including the effect of diffusioosmosis ($X_p$; Figure \ref{fig6}(b)), only the particles in scenario (I), where there is no pH change, move toward the dead-end of the pore. In the presence of NaOH in the pores (II-IV), particles leave the pore at early times (Figure \ref{fig6}(b), inset) due to the change in $\zeta_p$, which is opposite from the calculated trajectories in Figure \ref{fig1}(a) under a constant $\zeta_p$ assumption.

Finally, we add the influence of wall diffusioosmosis in the calculation of the 1D trajectories. Let $v_s$ be the diffusioosmotic velocity generated along the pore walls. Under the concentration gradient of multiple ions, $v_s$ has the form of 
\begin{equation}
v_s = -\frac{\epsilon}{\mu} \left(\frac{k_B T}{e} \right) \frac{\sum_{i}{D_i z_i \frac{\partial c_i}{\partial x}}}{\sum_{i}{D_i z_i^2 c_i}} \zeta_w - \frac{\epsilon}{8 \mu} \frac{\sum_{i}{z_i^2 \frac{\partial c_i}{\partial x}}}{\sum_{i}{z_i^2 c_i}} \zeta_w^2~ \label{eqslipv},
\end{equation} 
where $\zeta_w$ is the wall zeta potential \cite{gupta2019prf}. 

In a rectangular pore with dimensions $w$, $h$, and $\ell$, the fluid velocity generated along the pore due to the slip (diffusioosmotic) velocity $v_s$ is (see Appendix \ref{SectionAC} for details)  \cite{alessio2022jfm},
\begin{equation}
v_f = v_s \left[1 - \frac{6}{h^2}\left[1-\left(\frac{6}{\mathcal{W}}\right)\sum_{n=0}^{\infty}{\lambda_n^{-5}\tanh(\lambda_n \mathcal{W})}\right]^{-1} \left\{\left[\left(\frac{h}{2}\right)^2 - z^2 \right] - \sum_{n=0}^{\infty}{a_n \cos\left(\frac{\lambda_n z}{h/2}\right)\cosh\left(\frac{\lambda_n y}{h/2}\right)}\right\}\right]~,
\end{equation}
where $\mathcal{W}=w/h$, $\displaystyle{\lambda_n=\frac{2n+1}{2}\pi}$ (n=0,1,2,...), and $\displaystyle{a_n=\frac{h^2 (-1)^n}{\lambda_n^3 \cosh(\lambda_n \mathcal{W})}}$. For $w=100~\mu$m and $h=50~\mu$m (or $\mathcal{W}=2$), the fluid velocity along the centerline $(x,0,0)$ is equal to $-v_s$ (Figure \ref{figA2}). Therefore, for the analysis including diffusioosmosis, we calculate the centerline trajectory $x_c$, which follows
\begin{equation}
\frac{\partial x_c}{\partial t} = u_p(x_c,t)-v_s(x_c,t);~ x_c(t=0)=x_0~.
\end{equation}

The nondimensional centerline trajectory including diffusioosmosis ($X_c$) is calculated and plotted versus dimensional time in Figure \ref{fig6}(c). For the wall zeta potential $\zeta_w$, we used a linear fit (linear in pH) to the cited data  (Figure \ref{figA4}(a); see Appendix D for details) \cite{kirby2004-2}, which has a functional form $\zeta_w=-8(\ce{pH}-2)$ (mV). Only after including both the influences of wall diffusioosmosis and the varying $\zeta_p$ do we obtain the same trends between the compaction experiments (Figure \ref{fig5}(b-e)) and the trajectory calculations (Figure \ref{fig6}(c)). Also, the calculated centerline trajectory can be directly compared with the entrainment front (Figure \ref{figA1}(b); see Appendix \ref{SectionAC} for details). The calculated and measured $X_c$ in all four experiments (I-IV) show good agreement (Figure \ref{fig6}(c)). From the experiments and model calculations, we confirm that for a-PS, what looked like the behavior of negatively charged particles is in fact combined effects of varying $\zeta_p$ and wall diffusioosmosis-driven liquid flow in the pores.

After performing various systematic studies for diffusiophoresis along a pH gradient, we can finally provide explanations to the initial observation that motivated our paper: in the compaction configuration, a-PS particles appear to behave like negatively charged PS along the NaOH concentration gradient (Figure \ref{fig1}(d)). Zeta potential measurements and the particle patch experiments show that the a-PS particles do change their sign of $\zeta_p$ at high pH. However, compaction of a-PS particles under the NaOH concentration gradient does not happen only because of varying $\zeta_p$, but due to the strong influence of $\zeta_w$ and wall diffusioosmosis. The zeta potential of PDMS stays negative along a wide range of pH values, with small magnitudes at low pH and large magnitudes at high pH (Figure \ref{figA4}(a)). Therefore, the diffusioosmotic flow is fast enough in the presence of 10-20 mM NaOH in the pores (pH$_p\approx 12$) and make the particle behaviors appear to be that of negatively charged particles.  

\section{Diffusiophoresis of particles with different isoelectric points}
\label{SectionVI}

\begin{figure}[t!]
\centering
\includegraphics[width=\textwidth]{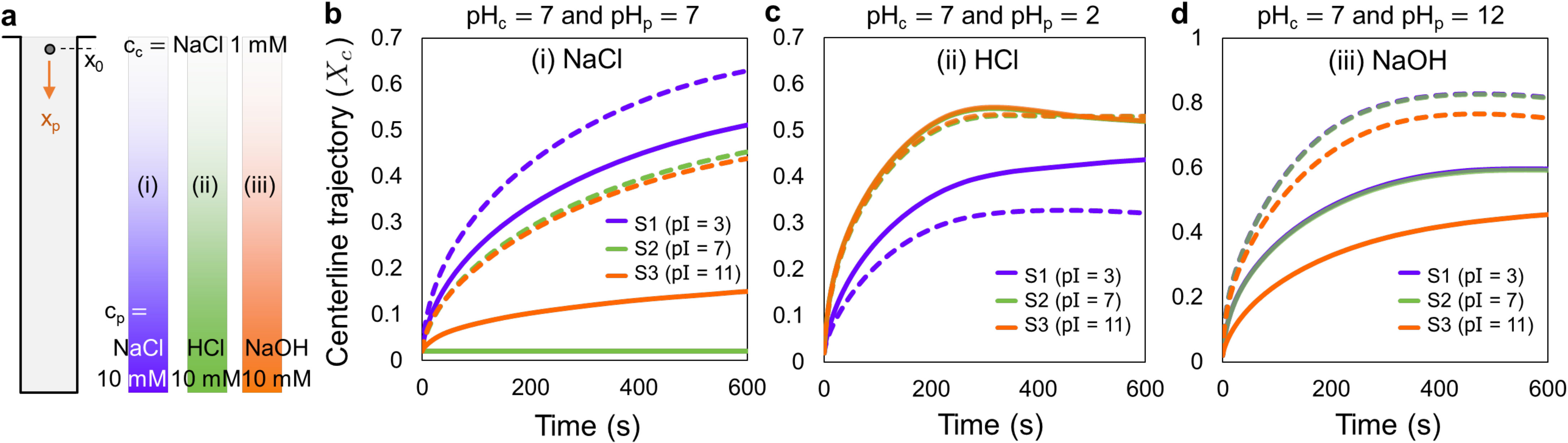}
\caption{\label{fig7} Model calculations for diffusiophoresis of particles with different isoelectric points under three different pH gradients. (a) Schematics showing the setup for calculation. We solve for centerline trajectories under the concentration gradients set by, respectively, (i) $c_c=1$ mM NaCl and $c_p=10$ mM NaCl (initially in the pore), (ii) $c_c=1$ mM NaCl and $c_p=10$ mM HCl, and (iii) $c_c=1$ mM NaCl and $c_c=10$ mM NaOH. (b-d) Calculated centerline trajectories for three sample particles S1 (pI $=3$), S2 (pI $=7$) and S3 (pI $=11$) are plotted versus time for (i-iii) with (dashed) and without (solid) the influence of wall diffusioosmosis. }
\end{figure}

The a-PS particles we used have the feature that the zeta potential changes sign at a high pH (pI $\approx 11.6$). The understandings we obtain from the diffusiophoresis (and diffusioosmosis) of a-PS particles can be applied to studies of other particles. For example, most proteins have their own isoelectric points \cite{gelelectrophoresis, proteinpurification}, so their diffusiophoretic behavior may look similar to that of a-PS particles reported here at their respective pI. To understand particle behaviors for a wide range of $\zeta_p(\ce{pH})$ in the presence of pH gradients, we performed more calculations using model particles with $\bar{\zeta}_p = -2 \tanh(0.5(\ce{pH}-\ce{pI}))$ (Figure \ref{figA4}(b)). For the isoelectric point, three values (\ce{pI=3, 7, 11}) are chosen to test diffusiophoretic behaviors under acidic and basic conditions. The functional form we have chosen does not represent any real particle, as every material shows different trends of $\zeta_p (\ce{pH})$ under pH gradient \cite{kirby2004-1,kirby2004-2}, but the basic features are plausible. In Figure \ref{fig7}(a), we show the three concentration gradients formed between (i) $c_c= 1$ mM and $c_p=10$ mM \ce{NaCl}, (ii) $c_c=1$ mM \ce{NaCl} and $c_p=10$ mM \ce{HCl}, and (iii) $c_c=1$ mM \ce{NaCl} and $c_p=10$ mM \ce{NaOH}, which represent (i) no \ce{pH} gradient, (ii) a \ce{pH} gradient between \ce{pH} $=2$ and 7, and (iii) between \ce{pH} $=7$ and 12, respectively. Sample particles S1 (\ce{pI} $= 3$), S2 (\ce{pI} $=7$), and S3 (\ce{pI} $=11$) (Figure \ref{figA4}(b)) are used for calculations.

The centerline trajectories ($X_c(\tau)$) in the presence and the absence of wall diffusioosmsosis are plotted versus dimensional time (Figure \ref{fig7}(b-d)). In the \ce{NaCl} concentration gradient (case i), all three sample particles (S1, S2, and S3) showed different diffusiophoretic behaviors, due to their differences in $\zeta_p$ at \ce{pH} $=7$. PDMS is negatively charged at pH $=7$, so the centerline flow velocity induced by diffusioosmosis is positive (flow toward dead-end). This additive effect of diffusioosmosis is observed for all three particles. Particle S2 (\ce{pI} $=7$) shows no diffusiophoresis at \ce{pH} $=7$, but in dead-end pore experiments, even the particles with $\zeta_p=0$ can move toward the dead-end due to the flow structure created inside the pore.  

When the pH gradient is formed between pH$_p=2$ and pH$_c=7$ ($c_p=10$ mM HCl and $c_c=1$ mM \ce{NaCl}), it can be guessed from the HCl mobility (equation 1; Figure \ref{fig1}(a)) that the particles with positive potential move toward the dead-end. The wall diffusioosmosis is directed inward (with negative $\zeta_w$), so the centerline flow velocity is negative (toward $x=0$). Of course, the details are more complex with multiple ions present in the pores. We obtain that the particles S2 (pI $=7$) and S3 (pI $=11$), which both have positive surface potentials between pH$_p$=2 and pH$_c$=7, move toward high HCl concentration (into the pore). In the cases with S2 and S3, diffusioosmosis makes little contribution to the centerline trajectory. The particle S1 (pI $=3$) undergoes a sign change in $\zeta_p$ between pH$_p=2$ and pH$_c=7$, and moves toward high HCl region at a slower speed than the other two samples. Diffusioosmosis along the wall induces a negative centerline flow velocity and the influence is strong for S1. 

Finally, when the pH gradient is formed between pH$_p=12$ and pH$_c=7$ ($c_p=10$ mM NaOH and $c_c=1$ mM NaCl), we observe that S1 (pI $=3$) and S2 (pI $=7$) move toward the dead-end of the pore as their $\zeta_p <0$ under pH $>7$. Similar to the experiments and calculations shown in Section \ref{SectionV} (Figures \ref{fig5} and \ref{fig6}), when a diffusioosmosis-driven centerline flow velocity is added to the particle motion, particles travel deeper into the pore. The sample S3 (pI $=11$) undergoes a sign change in $\zeta_p$ under the pH gradient, but the direction of motion is the same as othe two particles with a slower speed.

\section{Conclusions}
Our study of diffusiophoresis in the presence of a \ce{pH} gradient is motivated by the fact that natural and synthesized particles will have a zeta potential that changes with \ce{pH}. In particular, electrolyte-driven diffusiophoresis is characterized by the zeta potential of particles, but the zeta potential is not a fixed material property. Among several factors that influence the zeta potential of a particle surface we can list the ionic strength, \ce{pH}, solute type, etc., and here we highlighted the influence of \ce{pH}. 

The amine-modified polystyrene (a-PS) particles purchased from Sigma Aldrich have an isoelectric point \ce{pI} $\approx 11.6$, indicating that there is a sign change in the zeta potential between pH $=11$ and 12. Such extreme \ce{pH} can be set up by \ce{NaOH} of ionic strengths 1 and 10 mM, which is a common range used in many diffusiophoresis studies. In the compaction experiments of a-PS particles done in NaOH solutions (in the pore $c_p=10$ mM and main flow channel $c_c=1$ mM; \ce{pH}$_p=12$ and \ce{pH}$_c=11$), the a-PS particles, which are positively charged in moderate pH conditions, behaved like negatively charged PS particles. In order to understand the behavior of a-PS particles, we set up a charge regulation model to count the number of functional groups (\ce{SO4-} and \ce{NH3+}) that bind with \ce{H+} and obtain the zeta potential as a function of pH, i.e., $\zeta_p(\ce{pH})$. Then, using the function $\zeta_p(\ce{pH})$, we predicted the front propagation of particle patches in dead-end pores. When the influence of wall diffusioosmosis is negligible, the particle front propagation is well predicted by the 1D (centerline) trajectory. 

Compaction experiments that show a parabolic particle boundary require that the flow velocity driven by wall diffusioosmosis is included in the analyses for the motion of particles. By using various concentration gradients between pH $=7$ and pH $=12.3$ (Section \ref{SectionV}), compaction behaviors of a-PS particles in the dead-end pore were explained. From compaction experiments, it simply looked like that the a-PS particles are negatively charged in the dead-end pore, but the calculations and direct comparison with entrainment experiments showed that the response is the strong influence of wall diffusioosmosis that made the particles compact toward the dead-end. Since most surfaces have nonzero zeta potential when in contact with aqueous solutions, the influence of wall diffusioosmosis must be considered in the interpretation of particle motion under ion concentration gradients. 

The results from experimental and model studies on a-PS particles cannot be directly applied to other particles, as different particles have different $\zeta_p(\ce{pH})$. Nevertheless, we performed model calculations for sample particles with different pIs, and demonstrated that depending on how the concentration gradient is set up, diffusiophoretic behaviors of different particles vary. As diffusiophoresis of natural particles (cells, proteins, intracellular materials, etc.) in complex geometries (porous systems, varying configurations, confinements, etc.) is of interest to different research communities and in applications \cite{shim2022}, we believe that our results can provide the basis for further insights, especially when \ce{pH} gradients are present.


\begin{acknowledgments}
We thank the NSF for support via grant CBET-2127563.
\end{acknowledgments}

\appendix

\setcounter{figure}{0}
\makeatletter 
\renewcommand{\thefigure}{A\@arabic\c@figure}
\makeatother
\makeatletter 
\renewcommand{\thetable}{A\@arabic\c@table}
\makeatother

\section{Experimental methods}

\subsection{Particles used in the study}
Amine-modified polystyrene (a-PS; diameter = $1~\mu$m) particles are purchased from Sigma Aldrich (Product number: L9654). Two batches (MKCF6014 and MKCK7640) are used for compaction experiments (Sections \ref{SectionII} and \ref{SectionV}), zeta potentiometry (Figure \ref{fig2}(a)), and the particle patch experiments (Section \ref{SectionIII}). Polystyrene (PS; diameter = $1~\mu$m) particles are purchased from Thermo Fisher Scientific (Invitrogen, product number: F13082). One batch (2161862) is used for compaction experiments (Section \ref{SectionII}) and zeta potentiometry (Figure \ref{fig2}(a)). In order to avoid any quality change in the original particle suspension due to the storage conditions, all data are obtained within 1 week after the original seal is removed from the product. Since the surface of a-PS is positively charged for a wide range of pH (Figure \ref{fig2}(a)), a-PS particles sometimes adhered to the negatively charged PDMS walls. Such sticking behavior was observed frequently in the main channel where there is imposed flow, but rarely inside the pores. We observed that the adhesion of a-PS in the pores is significant only when the particles reside in the pore for a long time ($O(1)$ hr), and when the ionic strength is high ($O(100)$ mM), which acts to screen the surface charge. For the situations with high ionic strength, particles also lost their stability and agglomerated. Therefore, in the presented dead-end pore experiments performed at $c_i \approx 10$-20 mM, minor adhesion of a-PS particles on the PDMS walls does not affect our interpretation of the results.   

\subsection{Compaction and entrainment experiments}
\begin{figure}[h!]
\vspace{-1ex}
\centering
\includegraphics[width=0.7\textwidth]{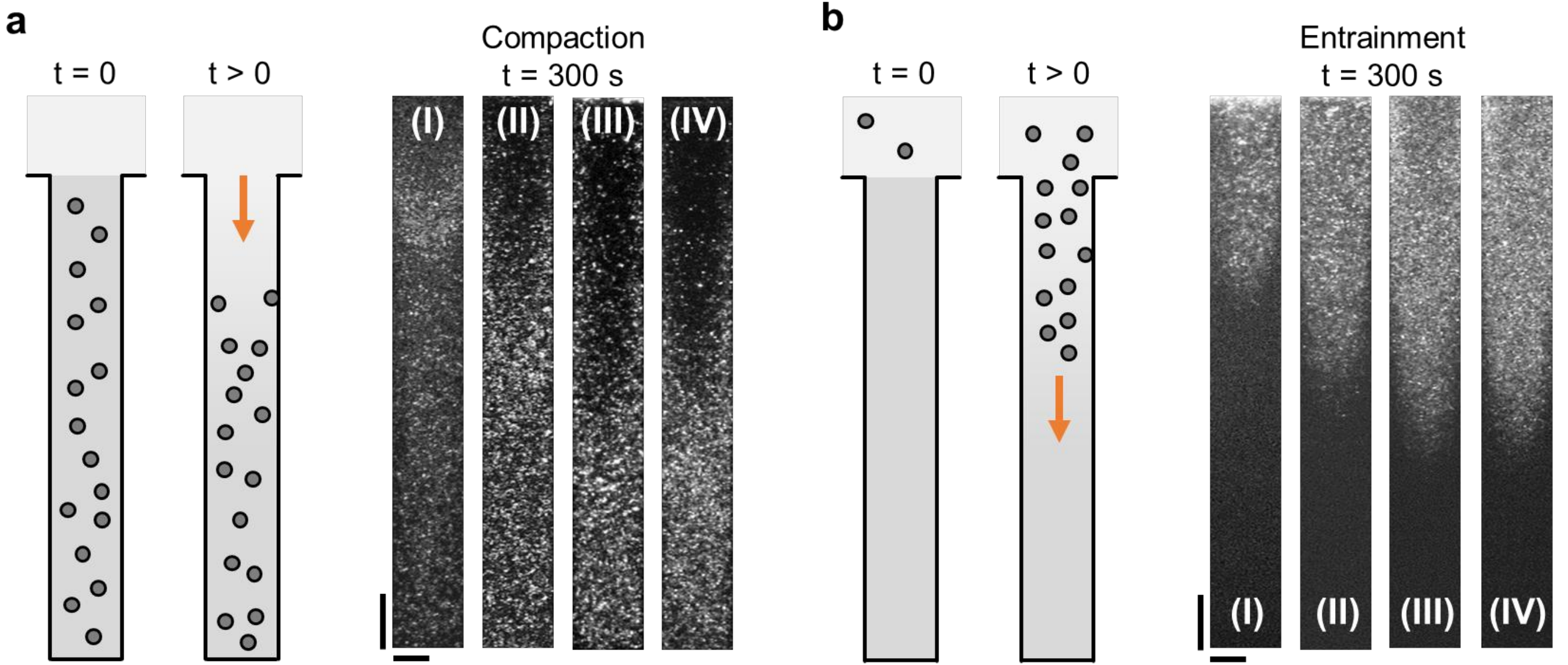}
\caption{\label{figA1} (a) Compaction and (b) entrainment experiments done under ion concentration gradients (I-IV) used in Figure \ref{fig5}. The experimental conditions (I-IV) are: (I) $c_c=2$ mM and $c_p=20$ mM NaCl, (II) $c_c=2$ mM NaCl and $c_p=10$ mM NaCl + 10 mM NaOH, (III) $c_c=2$ mM NaCl and $c_p=20$ mM NaOH, and (IV) $c_c=2$ mM NaOH and $c_p=20$ mM NaOH. }
\end{figure}

Compaction experiments shown in Sections \ref{SectionII} and \ref{SectionV} are performed with a microfluidic channel with five dead-end pores. The width, height, and length of the pores are, respectively, $w=100~\mu$m, $h=50~\mu$m, and $\ell=1$ mm. For the main channel, width, height, and the length are, respectively, $W=750~\mu$m, $H=150~\mu$m, and $\ell=5$ cm. The pores are initially filled with a particle suspension (0.02 $\%$v/v) at electrolyte concentration $c_p$. Then an air bubble, followed by the second liquid without any particles (concentration $c_c$) is flowed into the main channel at a mean flow speed $\langle u \rangle = 1$ mm/s. Immediately after the two liquids contact with each other, the flow speed in the main channel is decreased to $\langle u \rangle$ = 50 $\mu$m/s. The entrainment experiments shown in Figure \ref{figA1} are done with the same flow settings with the compaction experiments, except for the initial condition for particles. The a-PS particles are initially suspended in the second aqueous solution, and enter the pores after the two liquids contact with each other. Both compaction and entrainment of fluorescent a-PS particles in the pores are recorded with an inverted microscope (Leica DMI4000B) with 1 s interval. Fluorescent imaging is acquired using a 5x objective lens (Leica 506303; numerical aperture 0.12) and a Leica DFC360 FX camera. Detailed graphical descriptions of compaction and entrainment configurations can be found in \cite{wilson2020}. 

\subsection{Particle patch experiments}
The particle patch experiments shown in Section \ref{SectionIV} are performed with the same microfluidic channel used for compaction experiments. The pores are initially filled with 10 mM HCl solution, without any particles, then followed by the first air bubble as a spacer, after which the second liquid (10 mM HCl solution with suspended a-PS particles at 0.1 $\%$v/v) is flowed into the main channel at $\langle u \rangle = 1$ mm/s. The flow of particle suspension is maintained for 30 s, until the penetration of streamlines at the pore inlet forms a particle patch of the initial depth $\approx w$. Then, a second air bubble is introduced, followed by the third liquid (NaOH with $c_c$ = 1, 3, 5, and 10 mM), which is flowed in the main channel at $\langle u \rangle=1$ mm/s. Immediately after the liquid in the pores and the \ce{NaOH} solution in the main channel contact with each other, the flow speed in the main channel is decreased to $\langle u \rangle$ = 50 $\mu$m/s. Diffusiophoresis of the particle patches is recorded with an inverted microscope (Leica DMI4000B) with images taken at 1 s interval. Fluorescent imaging is acquired using a 5x objective lens (Leica 506303; numerical aperture 0.12) and a Leica DFC360 FX camera. A graphical description of the particle patch configuration can be found in \cite{alessio2022jfm}. 

To analyze the propagation of the particle front, a single image stack representing the five pores from each experiment is used. Image stacks of five pores are overlaid by adding the gray values using ImageJ, then the kymograph obtained from the combined stack (Figure \ref{fig3}(c)) is used to directly detect the front position at each time. Averages of four or five experiments are plotted in Figure \ref{fig4}(b), along with the error bars (standard deviation). 

\section{Charge regulation model for surface-modified polystyrene particles}

Consider a surface chemistry model that counts the acidic and basic functional groups, such as sulfate, carboxylate and amine groups. In this paper, we follow the charge regulation model discussed in \cite{healy1978ionizable}, which include (de)protonization of acidic and basic functional groups. Commercial microspheres may have complex surface structures depending on the manufacturers' production protocols. As we do not know all of the chemical details, we formulate the model with both known and unknown factors, and try to fit our zeta potential measurements using the unknowns as the fitting parameters. 

\subsection{Counting acidic and basic functional groups on particle surface}

\subsubsection{Acidic functional groups (--\ce{COO-}, --\ce{SO4-})}

The acidic functional groups follow the reaction 
\begin{equation}
    \text{HA} \leftrightharpoons \text{H}^+ + \text{A}^{-}~,
\end{equation}
and $\text{A}^-$ contributes to the negative surface charge of the particles. The acid dissociation constant \ce{K_A} is defined as 
\begin{equation}
    \ce{K_A}=\frac{\ce{[H+][A^-]}}{\ce{[HA]}}.
\end{equation}

Let \ce{n_A} the total number density of the acidic surface groups. Then we know that 
\begin{subequations}
\begin{align}
    &~~\ce{n_A}=\ce{[HA]}+\ce{[A^-]}\\
    \Rightarrow &~~\ce{n_A}-\ce{[A^-]}= \frac{\ce{[H+]}\ce{[A^-]}}{\ce{K_A}} \\
    \Rightarrow &~~\ce{[A^-]} = \frac{\ce{n_A}}{1+\ce{[H+]}/\ce{K_A}}~.
\end{align}
\end{subequations}
\ce{H+} ions in the surface region follow the Boltzmann distribution $\displaystyle{\ce{[H+]}=10^{-\ce{pH}} \exp\left(-\frac{e (\psi_s-\psi_{\infty})}{k_B T}\right)}$, where $e$, $\psi_s-\psi_{\infty}$, $k_B$, and $T$ are, respectively, the elementary charge, surface potential relative to the bulk ($=$ zeta potential $\zeta_p$), Boltzmann constant, and the absolute temperature. In our study, we consider the Gouy-Chapman model for the electrical double layer, so that the zeta potential is defined as the potential difference between the surface and bulk. Therefore, the acidic surface groups' contribution to the charge density ($q_A$) is 
\begin{equation}
   q_A= -e\ce{[A^-]} = \frac{-e \ce{n_A}}{1+10^{\ce{pK_A}-\ce{pH}}\exp \left(-\frac{e \zeta_p}{k_B T}\right)}~.
\end{equation}

\subsubsection{Basic functional groups (--\ce{NH2})}

As pH increases, \ce{H+} ions bind to the basic functional groups to make surfaces more positively charged. We consider the reaction 
\begin{equation}
    \ce{BH+} \leftrightharpoons \ce{B}+\ce{H+}
\end{equation}
and the acid dissociation constant of the conjugate acid \ce{BH+} 
\begin{equation}
    \ce{K_B} = \frac{\ce{[B][H+]}}{\ce{[BH+]}}~.
\end{equation}

Similar to the acidic functional groups, we can define \ce{n_B} as the total number of the basic surface groups. Then,
\begin{subequations}
\begin{align}
    &~~\ce{n_B}=\ce{[BH+]}+\ce{[B]}\\
    \Rightarrow&~~ \ce{[BH+]} = \frac{\ce{n_B[H+]/K_B}}{1+\ce{[H+]/K_B}} \\
    \therefore ~~q_B=e\ce{[BH+]} &=\frac{e \ce{n_B} 10^{\ce{pK_B-pH}}\exp\left(-\frac{e \zeta_p}{k_B T}\right)}{1+10^{\ce{pK_B-pH}}\exp\left(-\frac{e \zeta_p}{k_B T}\right)}~,
\end{align}
\end{subequations}
where $q_B$ is the positive charge density. 
 
According to the potential measurements and the technical notes distributed by the manufacturers \cite{sigma, thermofisher}, we can assume that our amine-modified polystyrene particles have amine and sulfate groups. Thus the surface charge density $q$ can be described as
\begin{equation}
    q = \frac{-e \ce{n_A}}{1+10^{\ce{pK_A}-\ce{pH}}\exp \left(-\frac{e \zeta_p}{k_B T}\right)}+\frac{e \ce{n_B} 10^{\ce{pK_B-pH}}\exp\left(-\frac{e \zeta_p}{k_B T}\right)}{1+10^{\ce{pK_B-pH}}\exp\left(-\frac{e \zeta_p}{k_B T}\right)}~ \label{chargeregulation}.
\end{equation}
Either sulfate or carboxylate functional groups can be used during the polymerization of styrene, and then the amine-modification step is applied to the surface. Therefore, final product can have multiple kinds of functional groups. Since zeta potential measurements of Sigma Aldrich particles (Figure 1(c)) show almost constant values for pH between 3 and 10, carboxylate functional groups (\ce{pK_A} $=5$) can be neglected. \ce{pK_A} of sulfate groups is 2, and \ce{pK_B} of \ce{NH3+} is not known for this specific product. Surface coverage of the functional groups is estimated as 30-300 \AA$^2$ per charge group \cite{sigma}. 

\vspace{-1ex}
\subsection{Zeta potential: comparison with the measurements}

Zeta potential measurements for a-PS particles are done in the presence of 10 mM NaCl as the background electrolyte. Except for the case of pH$=2$ and 12 (10 mM HCl and 10 mM NaOH, repsectively), the ionic strength is controlled by the background 10 mM NaCl. For all measurements, we can say that the thickness of electical double layer (EDL) is determined by the ionic strength $c=10$ mM. Therefore, it is reasonable to use the Gouy-Chapman formulation for a binary system to balance the zeta potential $\zeta_p$ and surface charge density $q$
\begin{equation}
    q=\frac{2 \epsilon k_B T}{\lambda_D e} \sinh\left(\frac{e \zeta_p}{2 k_B T} \right)=4 c e \lambda_D \sinh\left(\frac{e \zeta_p}{2 k_B T} \right)~\label{gcapp}. 
\end{equation}
$\epsilon$ is the dielectric permittivity (Table \ref{table1}), and the Debye length is defined as $\displaystyle{\lambda_D=\sqrt{\frac{\epsilon k_B T}{2 e^2 c}}}$, where $c=10$~mM.
We obtain the zeta potential as a function of pH by balancing the equation (\ref{chargeregulation}) and (\ref{gcapp}) with fitting parameters $e$\ce{n_A} $= 0.0402$ C/m$^2$, $e$\ce{n_B} $=0.0576$ C/m$^2$ and \ce{pK_B} $=12.1$. A least squares fit is used with $e$\ce{n_A} $=\pm0.0001$ C/m$^2$, $e$\ce{n_B} $=\pm 0.0001$ C/m$^2$, and \ce{pK_B} $=\pm0.05$, with a condition $\zeta_p(\ce{pH}=12)<-10$ mV. The solution is compared with the measured data in Figure \ref{fig2}(c). The number density $\ce{n_A}+\ce{n_B} = 6.11 \times 10^{17}$ per m$^2$ corresponds to 164  \AA$^2$ per charge group, and is consistent with the values given in the manufacturer's technical note \cite{sigma}.

\section{Diffusiophoresis and diffusioosmosis in the presence of pH gradient (Main text Sections \ref{SectionV}-\ref{SectionVI})} \label{SectionAC}
\subsection{Compaction and entrainment experiments}

For the four experimental conditions (I-IV) shown in Figure \ref{fig5}, we also performed a set of entrainment experiments (see methods for details). While keeping the ion concentration gradients consistent, and by suspending the particles in the second liquid (at $c_c$), we can entrain the a-PS particles into the pores that are initially filled with the first solution (at $c_p$, without particles). In this way, the fastest particles along the centerline can be visualized as the position of the entrainment front, which is measured at different times and plotted versus time in Figure \ref{fig6}(c). The calculated centerline trajectory $X_c$ and the measured entrainment front show good agreement for all four conditions (I-IV).

\subsection{Flow velocity in a dead-end pore generated by the wall diffusioosmosis}

\begin{figure}[t!]
\centering
\includegraphics[width=\textwidth]{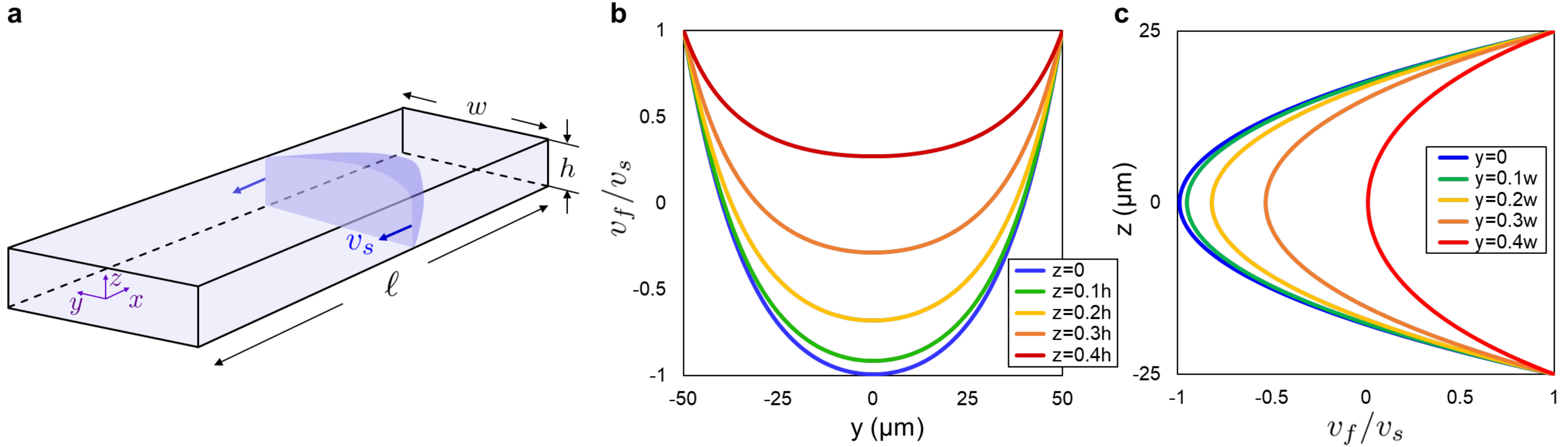}
\caption{\label{figA2} Flow velocity in a dead-end pore. (a) Schematic of a dead-end pore of the width, height and length, respectively, $w$, $h$, and $\ell$. (b,c) Velocity profiles plotted versus (b) $y$ and (c) $z$ for different values of $z$ and $y$, respectively. $w=100~\mu$m and $h=50~\mu$m are chosen to match with our experiments. Note that the centerline velocity is $-v_s$.}
\end{figure}

In a rectangular pore with width, height, and length, respectively, $w$, $h$, and $\ell$, the wall slip velocity $v_s(x)$ (equation \ref{eqslipv}) generated by diffusioosmosis induces liquid flow (Figure \ref{figA2}(a)) \cite{alessio2021prf, alessio2022jfm}. The flow velocity $v_f (x)$ can be obtained by applying lubrication approximation ($w\ll \ell$ and $h\ll \ell$). Thus we solve
\begin{align}
\frac{\partial^2 v_f}{\partial y^2}+\frac{\partial^2 v_f}{\partial z^2} - \frac{1}{\mu}\frac{dp}{dx} &=0~~;~~~v_f=v_s~~\text{at}~~y=\pm\frac{w}{2}~,~z=\pm\frac{h}{2}\\
\frac{\partial p}{\partial y}=\frac{\partial p}{\partial z} &=0\\
\frac{\partial v_f}{\partial x}+\frac{\partial v_{fy}}{\partial y}+\frac{\partial v_{fz}}{\partial z}&=0~.
\end{align}
We assume that the flow is only in the $x$-direction, and the assumption is reasonable as we do not include the inlet and dead-end region in the analysis. In the dead-end pore, the volumetric flow rate is zero, so that the fluid mass is conserved:
$\int_{0}^{\frac{h}{2}}\int_{0}^{\frac{w}{2}} ~v_f~ dydz = 0$

Equation (1) can be solved with separation of variables, and we obtain the well-known series solution that satisfies the boundary condition as 
\begin{equation}
v_f=-\frac{1}{2\mu}\frac{dp}{dx}\left\{\left[\left(\frac{h}{2}\right)^2-z^2\right]-\sum_{n=0}^{\infty} a_n \cos\left(\frac{\lambda_n z}{h/2}\right)\cosh\left(\frac{\lambda_n y}{h/2}\right)\right\}+v_s~,
\end{equation}
where $\displaystyle{a_n = \frac{h^2 (-1)^n}{\lambda_n^3 \cosh(\lambda_n w/h)}}$ and $\displaystyle{\lambda_n=\frac{2n+1}{2}\pi}$ ($n=0,1,2,...$).

\bigskip
Using the zero net flux condition, we obtain
\begin{equation}
Q=4\int_{0}^{\frac{h}{2}}\int_{0}^{\frac{w}{2}} ~v_f~ dydz = -\frac{w h^3}{12 \mu}\frac{dp}{dx}\left[1-\left(\frac{6}{\mathcal{W}}\right)\sum_{n=0}^\infty \lambda_n^{-5} \tanh(\lambda_n \mathcal{W})\right]+v_s w h = 0~,
\end{equation}
where $\mathcal{W}=w/h$. Therefore,
\begin{equation}
\frac{dp}{dx} = \frac{12 \mu v_s}{h^2}\left[1-\left(\frac{6}{\mathcal{W}}\right)\sum_{n=0}^\infty \lambda_n^{-5} \tanh(\lambda_n \mathcal{W})\right]^{-1}~ \label{eqdpdx}.
\end{equation}

Rewriting the flow velocity $v_f$ using equation (\ref{eqdpdx}), we get
\begin{equation}
v_f = v_s\left[1-\frac{6}{h^2}\left[1-\left(\frac{6}{\mathcal{W}}\right)\sum_{n=0}^\infty \lambda_n^{-5} \tanh(\lambda_n \mathcal{W})\right]^{-1} \left\{\left[\left(\frac{h}{2}\right)^2-z^2\right]-\sum_{n=0}^{\infty} a_n \cos\left(\frac{\lambda_n z}{h/2}\right)\cosh\left(\frac{\lambda_n y}{h/2}\right)\right\}\right].
\end{equation}

Note that $v_s(x)$ is the only function of $x$, and the flow velocity $v_f/v_s$ is plotted versus $y$ and $z$ in Figure \ref{figA2}(b,c). We obtain that the centerline velocity in the pore with $w=100~\mu$m and $h=50~\mu$m (or $\mathcal{W}=2$) is $-v_s$, and thus we obtain the 1D trajectory $x_c$ using $u_p - v_s$ as the centerline velocity. 

\bigskip

\subsection{Coupled diffusion of ions in the systems in Sections \ref{SectionV} and \ref{SectionVI}}

In the calculations for the particle front trajectory (Section \ref{SectionIV}), we considered the full Nernst-Planck equation for ion transport including the water reaction term. In Sections \ref{SectionV} and \ref{SectionIV}, we assume that the fast reaction $\ce{H2O}(\ell) \leftrightharpoons \ce{H+}+\ce{OH-}$ is in equilibrium for all time. Then, for the systems with initially high NaOH concentration in the pores ($c_p \geq 10$ mM; Scenarios (II-IV) in Section \ref{SectionV} and (iii) in Section \ref{SectionVI}), we solve

\begin{subequations}
\begin{align}
   & \frac{\partial c_{\scaleto{\ce{Na}}{4pt}}}{\partial t} = D_{\scaleto{\ce{Na}}{4pt}} \frac{\partial^2 c_{\scaleto{\ce{Na}}{4pt}}}{\partial x^2} + \frac{D_{\scaleto{\ce{Na}}{4pt}}e}{k_B T} \left( \frac{\partial c_{\scaleto{\ce{Na}}{4pt}}}{\partial x} \frac{\partial \psi}{\partial  x} + c_{\scaleto{\ce{Na}}{4pt}}\frac{\partial^2 \psi}{\partial x^2}\right) \label{eqappcna}\\
   & \frac{\partial c_{\scaleto{\ce{Cl}}{4pt}}}{\partial t} = D_{\scaleto{\ce{Cl}}{4pt}}\frac{\partial^2 c_{\scaleto{\ce{Cl}}{4pt}}}{\partial x^2} - \frac{D_{\scaleto{\ce{Cl}}{4pt}} e}{k_B T} \left( \frac{\partial c_{\scaleto{\ce{Cl}}{4pt}}}{\partial x} \frac{\partial \psi}{\partial x} + c_{\scaleto{\ce{Cl}}{4pt}}\frac{\partial^2 \psi}{\partial x^2} \right) \label{eqappccl}\\
  &  \frac{\partial c_{\scaleto{\ce{OH}}{4pt}}}{\partial t} = D_{\scaleto{\ce{OH}}{4pt}}\frac{\partial^2 c_{\scaleto{\ce{OH}}{4pt}}}{\partial x^2} - \frac{D_{\scaleto{\ce{OH}}{4pt}}e}{k_B T} \left( \frac{\partial c_{\scaleto{\ce{OH}}{4pt}}}{\partial x} \frac{\partial \psi}{\partial  x} + c_{\scaleto{\ce{OH}}{4pt}}\frac{\partial^2 \psi}{\partial x^2}\right) ~\label{eqappcoh}\\
   & c_{\scaleto{\ce{H}}{4pt}}=K_w/c_{\scaleto{\ce{OH}}{4pt}}~,
\end{align}
\end{subequations}
along with the electroneutrality and zero current condition (equations 9 and 10). For the case with initially high \ce{HCl} concentration in the pore, ($c_p=10$ mM HCl and $c_c=1$ mM NaCl; Scenario (ii) in Section \ref{SectionVI}), we solve equations (\ref{eqappcna}), (\ref{eqappccl}), and
\begin{subequations}
\begin{align}
 &\frac{\partial c_{\scaleto{\ce{H}}{4pt}}}{\partial t} = D_{\scaleto{\ce{H}}{4pt}}\frac{\partial^2 c_{\scaleto{\ce{H}}{4pt}}}{\partial x^2} - \frac{D_{\scaleto{\ce{H}}{4pt}}e}{k_B T} \left( \frac{\partial c_{\scaleto{\ce{H}}{4pt}}}{\partial x} \frac{\partial \psi}{\partial  x} + c_{\scaleto{\ce{H}}{4pt}}\frac{\partial^2 \psi}{\partial x^2}\right)~\\
 &c_{\scaleto{\ce{OH}}{4pt}}=K_w/c_{\scaleto{\ce{H}}{4pt}}~,
 \end{align}
 \end{subequations}
along with the electroneutrality and zero current condition (equations 9 and 10).

\begin{figure}[t!]
\centering
\includegraphics[width=0.35\textwidth]{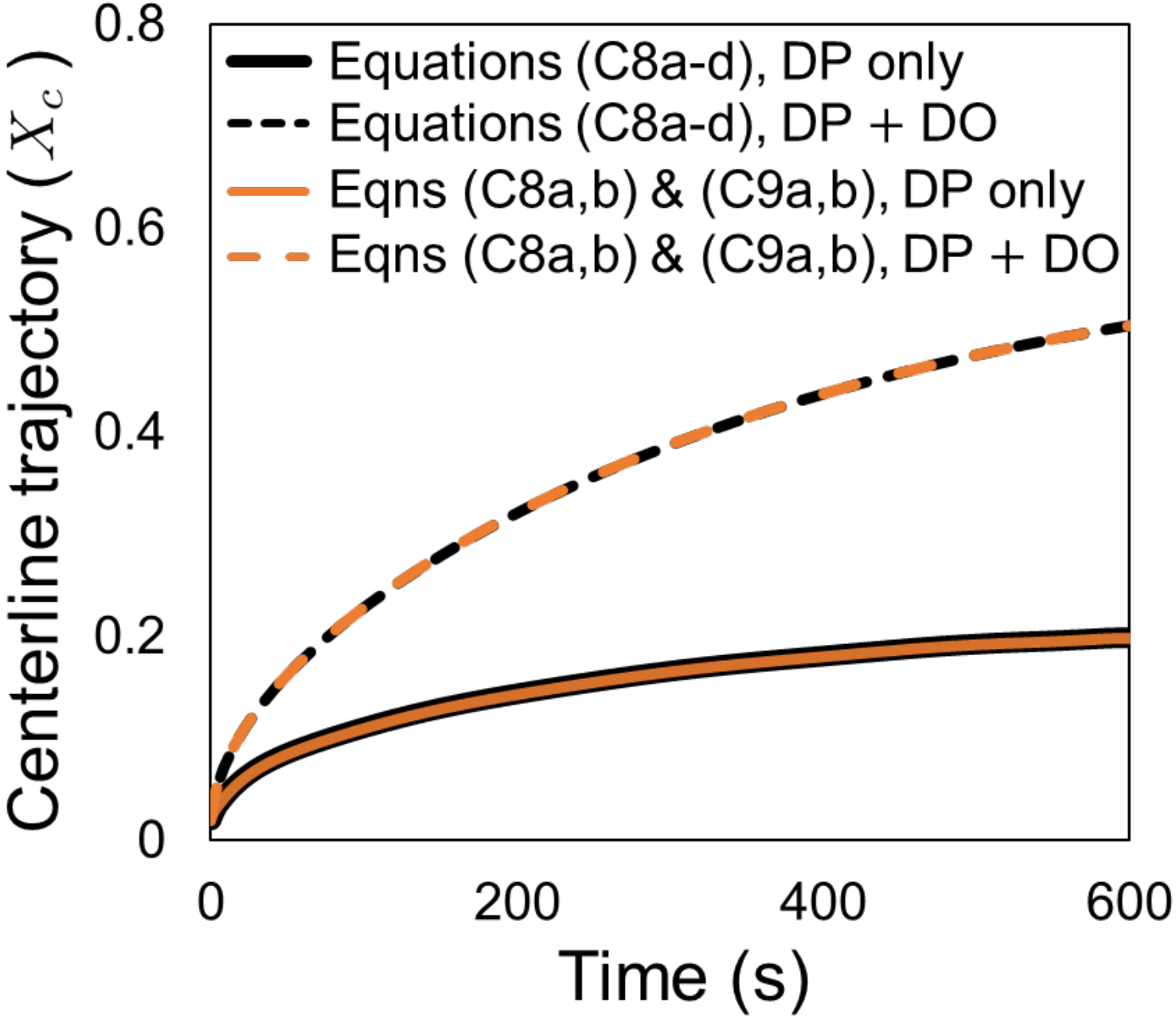}
\caption{\label{figA3} Calculations for centerline trajectories ($X_c$) in the absence and the presence of diffusioosmosis (DO) for the case (I) in Section \ref{SectionV} ($c_p=20$ mM NaCl and $c_c=2$ mM NaCl). The 1D diffusiophoretic (DP) trajectories in the absence and the presence of DO are compared between two sets of ion transport equations: equations (C8a-d), and (C8a,b) and (C9a,b). We obtain the same results from both ways of calculating the ion transport. }
\vspace{-1ex}
\end{figure}

Either approach can be applied for the cases with no pH gradient (NaCl only; Scenarios (I) in Section \ref{SectionV} and (i) in Section \ref{SectionVI}), and we confirm that the solutions obtained from both sets of equations are the same (Figure \ref{figA3}). 

The trajectory calculations ($X_p$ and $X_c$) are done with the nondimensional equations for Section \ref{SectionV} and \ref{SectionVI} using MATLAB routines and employing a central difference scheme. For $0<X<1$ and $0<\tau<5.6$, the spatial and time steps were chosen as, respectively, $\delta X=0.02$ and $\delta \tau = 10^{-6}$.

\subsection{Zeta potential of PDMS ($\zeta_w (\ce{pH})$) and sample particles used for Section \ref{SectionVI} }

\begin{figure}[h!]
\centering
\includegraphics[width=0.7\textwidth]{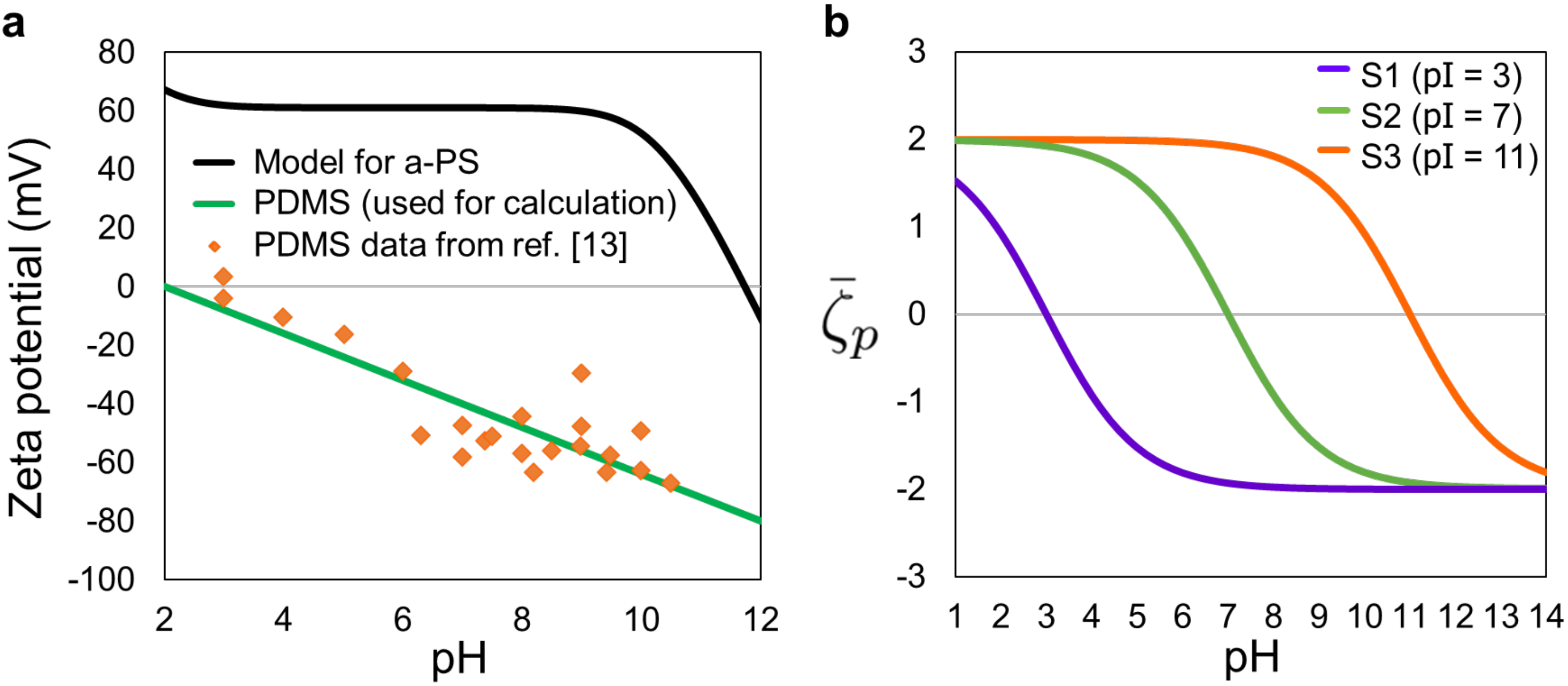}
\caption{\label{figA4} Zeta potential values used for calculations. (a) For the calculations presented in Section V, solutions to the equation (\ref{eqncr}, \ref{eqngc}) are used for particle (a-PS) zeta potential. For the wall zeta potential, we use the linear fit (linear in pH; green line) to the cited data ($\zeta_w=-8(\ce{pH}-2)$ (mV)). (b) For the calculations presented in Secton VI, we use three different sample particles (S1, S2, and S3), which have the zeta potential in a functional form of $\bar{\zeta}_p=-2\tanh(0.5 (\ce{pH}-\ce{pI}))$ and three different isoelectric points (pI=3,7,11).    }
\end{figure}

The wall zeta potential used for the diffusioosmosis calculations (equation \ref{eqslipv}) is obtained by applying a linear fit (linear in pH) to the data cited from \cite{kirby2004-2} (Figure \ref{figA4}(a)), and the functional form is $\zeta_w=-8(\ce{pH}-2)$ (mV). For model calculations reported in Section \ref{SectionVI}, three sample particles (S1, S2 and S3) are chosen so that particles with different isoelectric points can be examined for their diffusiophoretic behaviors. $\bar{\zeta}_p = -2 \tanh(0.5(\ce{pH}-\ce{pI}))$, and the values of pI are, 3, 7 and 11, respectively for three particles. The functional form $\bar{\zeta}_p = -2 \tanh(0.5(\ce{pH}-\ce{pI}))$ does not represent any real particle and is chosen for convenience to illustrate the role of of the difference of \ce{pH} and \ce{pI} on the particle motion.

\pagebreak


\bibliography{pHDiffusiophoresis}

\end{document}